\documentclass[journal,twoside,twocolumn]{IEEEtran}
\usepackage{cite}
\usepackage[T1]{fontenc}
\usepackage{amssymb}
\usepackage{amsmath}
\usepackage{paralist}
\usepackage[font=footnotesize]{caption}
\usepackage{subcaption}
\usepackage{graphicx}
\usepackage{booktabs} %\toprule \midrule \bottomrule
\usepackage{algorithm}
\usepackage{amsthm}
\usepackage{multirow}
\usepackage{microtype} % best command ever
\usepackage{tablefootnote}
\usepackage{scrextend}
\usepackage{balance}
\usepackage{algpseudocode}

\usepackage{amssymb}
\usepackage{amsfonts}
\usepackage{mathrsfs}
\usepackage{xspace}
\usepackage{bm}
\usepackage{upgreek}

\newcommand{\safemath}[2]{\newcommand{#1}{\ensuremath{#2}\xspace}}

\safemath{\bma}{\mathbf{a}}
\safemath{\bmb}{\mathbf{b}}
\safemath{\bmc}{\mathbf{c}}
\safemath{\bmd}{\mathbf{d}}
\safemath{\bme}{\mathbf{e}}
\safemath{\bmf}{\mathbf{f}}
\safemath{\bmg}{\mathbf{g}}
\safemath{\bmh}{\mathbf{h}}
\safemath{\bmi}{\mathbf{i}}
\safemath{\bmj}{\mathbf{j}}
\safemath{\bmk}{\mathbf{k}}
\safemath{\bml}{\mathbf{l}}
\safemath{\bmm}{\mathbf{m}}
\safemath{\bmn}{\mathbf{n}}
\safemath{\bmo}{\mathbf{o}}
\safemath{\bmp}{\mathbf{p}}
\safemath{\bmq}{\mathbf{q}}
\safemath{\bmr}{\mathbf{r}}
\safemath{\bms}{\mathbf{s}}
\safemath{\bmt}{\mathbf{t}}
\safemath{\bmu}{\mathbf{u}}
\safemath{\bmv}{\mathbf{v}}
\safemath{\bmw}{\mathbf{w}}
\safemath{\bmx}{\mathbf{x}}
\safemath{\bmy}{\mathbf{y}}
\safemath{\bmz}{\mathbf{z}}
\safemath{\bmzero}{\mathbf{0}}
\safemath{\bmone}{\mathbf{1}}

\bmdefine{\biad}{a}
\bmdefine{\bibd}{b}
\bmdefine{\bicd}{c}
\bmdefine{\bidd}{d}
\bmdefine{\bied}{e}
\bmdefine{\bifd}{f}
\bmdefine{\bigd}{g}
\bmdefine{\bihd}{h}
\bmdefine{\biid}{i}
\bmdefine{\bijd}{j}
\bmdefine{\bikd}{k}
\bmdefine{\bild}{l}
\bmdefine{\bimd}{m}
\bmdefine{\bind}{n}
\bmdefine{\biod}{o}
\bmdefine{\bipd}{p}
\bmdefine{\biqd}{q}
\bmdefine{\bird}{r}
\bmdefine{\bisd}{s}
\bmdefine{\bitd}{t}
\bmdefine{\biud}{u}
\bmdefine{\bivd}{v}
\bmdefine{\biwd}{w}
\bmdefine{\bixd}{x}
\bmdefine{\biyd}{y}
\bmdefine{\bizd}{z}

\bmdefine{\bixid}{\xi}
\bmdefine{\bilambdad}{\lambda}
\bmdefine{\bimud}{\mu}
\bmdefine{\bithetad}{\theta}
\bmdefine{\biphid}{\phi}
\bmdefine{\bideltad}{\delta}

\safemath{\bmia}{\biad}
\safemath{\bmib}{\bibd}
\safemath{\bmic}{\bicd}
\safemath{\bmid}{\bidd}
\safemath{\bmie}{\bied}
\safemath{\bmif}{\bifd}
\safemath{\bmig}{\bigd}
\safemath{\bmih}{\bihd}
\safemath{\bmii}{\biid}
\safemath{\bmij}{\bijd}
\safemath{\bmik}{\bikd}
\safemath{\bmil}{\bild}
\safemath{\bmim}{\bimd}
\safemath{\bmin}{\bind}
\safemath{\bmio}{\biod}
\safemath{\bmip}{\bipd}
\safemath{\bmiq}{\biqd}
\safemath{\bmir}{\bird}
\safemath{\bmis}{\bisd}
\safemath{\bmit}{\bitd}
\safemath{\bmiu}{\biud}
\safemath{\bmiv}{\bivd}
\safemath{\bmiw}{\biwd}
\safemath{\bmix}{\bixd}
\safemath{\bmiy}{\biyd}
\safemath{\bmiz}{\bizd}

\safemath{\bmxi}{\bixid}
\safemath{\bmlambda}{\bilambdad}
\safemath{\bmmu}{\bimud}
\safemath{\bmtheta}{\bithetad}
\safemath{\bmphi}{\biphid}
\safemath{\bmdelta}{\bideltad}

\safemath{\bA}{\mathbf{A}}
\safemath{\bB}{\mathbf{B}}
\safemath{\bC}{\mathbf{C}}
\safemath{\bD}{\mathbf{D}}
\safemath{\bE}{\mathbf{E}}
\safemath{\bF}{\mathbf{F}}
\safemath{\bG}{\mathbf{G}}
\safemath{\bH}{\mathbf{H}}
\safemath{\bI}{\mathbf{I}}
\safemath{\bJ}{\mathbf{J}}
\safemath{\bK}{\mathbf{K}}
\safemath{\bL}{\mathbf{L}}
\safemath{\bM}{\mathbf{M}}
\safemath{\bN}{\mathbf{N}}
\safemath{\bO}{\mathbf{O}}
\safemath{\bP}{\mathbf{P}}
\safemath{\bQ}{\mathbf{Q}}
\safemath{\bR}{\mathbf{R}}
\safemath{\bS}{\mathbf{S}}
\safemath{\bT}{\mathbf{T}}
\safemath{\bU}{\mathbf{U}}
\safemath{\bV}{\mathbf{V}}
\safemath{\bW}{\mathbf{W}}
\safemath{\bX}{\mathbf{X}}
\safemath{\bY}{\mathbf{Y}}
\safemath{\bZ}{\mathbf{Z}}

\safemath{\bZero}{\mathbf{0}}
\safemath{\bOne}{\mathbf{1}}
\safemath{\bDelta}{\mathbf{\Delta}}
\safemath{\bLambda}{\mathbf{\UpLambda}}
\safemath{\bPhi}{\mathbf{\Upphi}}
\safemath{\bSigma}{\mathbf{\Upsigma}}
\safemath{\bOmega}{\mathbf{\Upomega}}
\safemath{\bTheta}{\mathbf{\Uptheta}}

\bmdefine{\biAd}{A}
\bmdefine{\biBd}{B}
\bmdefine{\biCd}{C}
\bmdefine{\biDd}{D}
\bmdefine{\biEd}{E}
\bmdefine{\biFd}{F}
\bmdefine{\biGd}{G}
\bmdefine{\biHd}{H}
\bmdefine{\biId}{I}
\bmdefine{\biJd}{J}
\bmdefine{\biKd}{K}
\bmdefine{\biLd}{L}
\bmdefine{\biMd}{M}
\bmdefine{\biOd}{N}
\bmdefine{\biPd}{O}
\bmdefine{\biQd}{P}
\bmdefine{\biRd}{R}
\bmdefine{\biSd}{S}
\bmdefine{\biTd}{T}
\bmdefine{\biUd}{U}
\bmdefine{\biVd}{V}
\bmdefine{\biWd}{W}
\bmdefine{\biXd}{X}
\bmdefine{\biYd}{Y}
\bmdefine{\biZd}{Z}

\bmdefine{\biDelta}{\Delta}
\bmdefine{\biLambda}{\Lambda}
\bmdefine{\biPhi}{\Phi}
\bmdefine{\biSigma}{\Sigma}
\bmdefine{\biOmega}{\Omega}
\bmdefine{\biTheta}{\Theta}

\safemath{\bimA}{\biAd}
\safemath{\bimB}{\biBd}
\safemath{\bimC}{\biCd}
\safemath{\bimD}{\biDd}
\safemath{\bimE}{\biEd}
\safemath{\bimF}{\biFd}
\safemath{\bimG}{\biGd}
\safemath{\bimH}{\biHd}
\safemath{\bimI}{\biId}
\safemath{\bimJ}{\biJd}
\safemath{\bimK}{\biKd}
\safemath{\bimL}{\biLd}
\safemath{\bimM}{\biMd}
\safemath{\bimN}{\biNd}
\safemath{\bimO}{\biOd}
\safemath{\bimP}{\biPd}
\safemath{\bimQ}{\biQd}
\safemath{\bimR}{\biRd}
\safemath{\bimS}{\biSd}
\safemath{\bimT}{\biTd}
\safemath{\bimU}{\biUd}
\safemath{\bimV}{\biVd}
\safemath{\bimW}{\biWd}
\safemath{\bimX}{\biXd}
\safemath{\bimY}{\biYd}
\safemath{\bimZ}{\biZd}

\safemath{\bimDelta}{\biDelta}
\safemath{\bimLambda}{\biLambda}
\safemath{\bimPhi}{\biPhi}
\safemath{\bimSigma}{\biSigma}
\safemath{\bimOmega}{\biOmega}
\safemath{\bimTheta}{\biTheta}

\safemath{\setA}{\mathcal{A}}
\safemath{\setB}{\mathcal{B}}
\safemath{\setC}{\mathcal{C}}
\safemath{\setD}{\mathcal{D}}
\safemath{\setE}{\mathcal{E}}
\safemath{\setF}{\mathcal{F}}
\safemath{\setG}{\mathcal{G}}
\safemath{\setH}{\mathcal{H}}
\safemath{\setI}{\mathcal{I}}
\safemath{\setJ}{\mathcal{J}}
\safemath{\setK}{\mathcal{K}}
\safemath{\setL}{\mathcal{L}}
\safemath{\setM}{\mathcal{M}}
\safemath{\setN}{\mathcal{N}}
\safemath{\setO}{\mathcal{O}}
\safemath{\setP}{\mathcal{P}}
\safemath{\setQ}{\mathcal{Q}}
\safemath{\setR}{\mathcal{R}}
\safemath{\setS}{\mathcal{S}}
\safemath{\setT}{\mathcal{T}}
\safemath{\setU}{\mathcal{U}}
\safemath{\setV}{\mathcal{V}}
\safemath{\setW}{\mathcal{W}}
\safemath{\setX}{\mathcal{X}}
\safemath{\setY}{\mathcal{Y}}
\safemath{\setZ}{\mathcal{Z}}
\safemath{\emptySet}{\varnothing}

\safemath{\colA}{\mathscr{A}}
\safemath{\colB}{\mathscr{B}}
\safemath{\colC}{\mathscr{C}}
\safemath{\colD}{\mathscr{D}}
\safemath{\colE}{\mathscr{E}}
\safemath{\colF}{\mathscr{F}}
\safemath{\colG}{\mathscr{G}}
\safemath{\colH}{\mathscr{H}}
\safemath{\colI}{\mathscr{I}}
\safemath{\colJ}{\mathscr{J}}
\safemath{\colK}{\mathscr{K}}
\safemath{\colL}{\mathscr{L}}
\safemath{\colM}{\mathscr{M}}
\safemath{\colN}{\mathscr{N}}
\safemath{\colO}{\mathscr{O}}
\safemath{\colP}{\mathscr{P}}
\safemath{\colQ}{\mathscr{Q}}
\safemath{\colR}{\mathscr{R}}
\safemath{\colS}{\mathscr{S}}
\safemath{\colT}{\mathscr{T}}
\safemath{\colU}{\mathscr{U}}
\safemath{\colV}{\mathscr{V}}
\safemath{\colW}{\mathscr{W}}
\safemath{\colX}{\mathscr{X}}
\safemath{\colY}{\mathscr{Y}}
\safemath{\colZ}{\mathscr{Z}}

\safemath{\opA}{\mathbb{A}}
\safemath{\opB}{\mathbb{B}}
\safemath{\opC}{\mathbb{C}}
\safemath{\opD}{\mathbb{D}}
\safemath{\opE}{\mathbb{E}}
\safemath{\opF}{\mathbb{F}}
\safemath{\opG}{\mathbb{G}}
\safemath{\opH}{\mathbb{H}}
\safemath{\opI}{\mathbb{I}}
\safemath{\opJ}{\mathbb{J}}
\safemath{\opK}{\mathbb{K}}
\safemath{\opL}{\mathbb{L}}
\safemath{\opM}{\mathbb{M}}
\safemath{\opN}{\mathbb{N}}
\safemath{\opO}{\mathbb{O}}
\safemath{\opP}{\mathbb{P}}
\safemath{\opQ}{\mathbb{Q}}
\safemath{\opR}{\mathbb{R}}
\safemath{\opS}{\mathbb{S}}
\safemath{\opT}{\mathbb{T}}
\safemath{\opU}{\mathbb{U}}
\safemath{\opV}{\mathbb{V}}
\safemath{\opW}{\mathbb{W}}
\safemath{\opX}{\mathbb{X}}
\safemath{\opY}{\mathbb{Y}}
\safemath{\opZ}{\mathbb{Z}}
\safemath{\opZero}{\mathbb{O}}
\safemath{\identityop}{\opI}

\safemath{\veca}{\bma}
\safemath{\vecb}{\bmb}
\safemath{\vecc}{\bmc}
\safemath{\vecd}{\bmd}
\safemath{\vece}{\bme}
\safemath{\vecf}{\bmf}
\safemath{\vecg}{\bmg}
\safemath{\vech}{\bmh}
\safemath{\veci}{\bmi}
\safemath{\vecj}{\bmj}
\safemath{\veck}{\bmk}
\safemath{\vecl}{\bml}
\safemath{\vecm}{\bmm}
\safemath{\vecn}{\bmn}
\safemath{\veco}{\bmo}
\safemath{\vecp}{\bmp}
\safemath{\vecq}{\bmq}
\safemath{\vecr}{\bmr}
\safemath{\vecs}{\bms}
\safemath{\vect}{\bmt}
\safemath{\vecu}{\bmu}
\safemath{\vecv}{\bmv}
\safemath{\vecw}{\bmw}
\safemath{\vecx}{\bmx}
\safemath{\vecy}{\bmy}
\safemath{\vecz}{\bmz}

\safemath{\veczero}{\bmzero}
\safemath{\vecone}{\bmone}
\safemath{\vecxi}{\bmxi}
\safemath{\veclambda}{\bmlambda}
\safemath{\vecmu}{\bmmu}
\safemath{\vectheta}{\bmtheta}
\safemath{\vecphi}{\bmphi}
\safemath{\vecdelta}{\bmdelta}

\safemath{\matA}{\bA}
\safemath{\matB}{\bB}
\safemath{\matC}{\bC}
\safemath{\matD}{\bD}
\safemath{\matE}{\bE}
\safemath{\matF}{\bF}
\safemath{\matG}{\bG}
\safemath{\matH}{\bH}
\safemath{\matI}{\bI}
\safemath{\matJ}{\bJ}
\safemath{\matK}{\bK}
\safemath{\matL}{\bL}
\safemath{\matM}{\bM}
\safemath{\matN}{\bN}
\safemath{\matO}{\bO}
\safemath{\matP}{\bP}
\safemath{\matQ}{\bQ}
\safemath{\matR}{\bR}
\safemath{\matS}{\bS}
\safemath{\matT}{\bT}
\safemath{\matU}{\bU}
\safemath{\matV}{\bV}
\safemath{\matW}{\bW}
\safemath{\matX}{\bX}
\safemath{\matY}{\bY}
\safemath{\matZ}{\bZ}
\safemath{\matzero}{\bmzero}

\safemath{\matDelta}{\bDelta}
\safemath{\matLambda}{\bLambda}
\safemath{\matPhi}{\bPhi}
\safemath{\matSigma}{\bSigma}
\safemath{\matOmega}{\bOmega}
\safemath{\matTheta}{\bTheta}

\safemath{\matidentity}{\matI}
\safemath{\matone}{\matO}

\safemath{\rnda}{A}
\safemath{\rndb}{B}
\safemath{\rndc}{C}
\safemath{\rndd}{D}
\safemath{\rnde}{E}
\safemath{\rndf}{F}
\safemath{\rndg}{G}
\safemath{\rndh}{H}
\safemath{\rndi}{I}
\safemath{\rndj}{J}
\safemath{\rndk}{K}
\safemath{\rndl}{L}
\safemath{\rndm}{M}
\safemath{\rndn}{N}
\safemath{\rndo}{O}
\safemath{\rndp}{P}
\safemath{\rndq}{Q}
\safemath{\rndr}{R}
\safemath{\rnds}{S}
\safemath{\rndt}{T}
\safemath{\rndu}{U}
\safemath{\rndv}{V}
\safemath{\rndw}{W}
\safemath{\rndx}{X}
\safemath{\rndy}{Y}
\safemath{\rndz}{Z}

\safemath{\rveca}{\bimA}
\safemath{\rvecb}{\bimB}
\safemath{\rvecc}{\bimC}
\safemath{\rvecd}{\bimD}
\safemath{\rvece}{\bimE}
\safemath{\rvecf}{\bimF}
\safemath{\rvecg}{\bimG}
\safemath{\rvech}{\bimH}
\safemath{\rveci}{\bimI}
\safemath{\rvecj}{\bimJ}
\safemath{\rveck}{\bimK}
\safemath{\rvecl}{\bimL}
\safemath{\rvecm}{\bimM}
\safemath{\rvecn}{\bimN}
\safemath{\rveco}{\bomO}
\safemath{\rvecp}{\bimP}
\safemath{\rvecq}{\bimQ}
\safemath{\rvecr}{\bimR}
\safemath{\rvecs}{\bimS}
\safemath{\rvect}{\bimT}
\safemath{\rvecu}{\bimU}
\safemath{\rvecv}{\bimV}
\safemath{\rvecw}{\bimW}
\safemath{\rvecx}{\bimX}
\safemath{\rvecy}{\bimY}
\safemath{\rvecz}{\bimZ}

\safemath{\rvecxi}{\bmxi}
\safemath{\rveclambda}{\bmlambda}
\safemath{\rvecmu}{\bmmu}
\safemath{\rvectheta}{\bmtheta}
\safemath{\rvecphi}{\bmphi}

\safemath{\rmatA}{\bimA}
\safemath{\rmatB}{\bimB}
\safemath{\rmatC}{\bimC}
\safemath{\rmatD}{\bimD}
\safemath{\rmatE}{\bimE}
\safemath{\rmatF}{\bimF}
\safemath{\rmatG}{\bimG}
\safemath{\rmatH}{\bimH}
\safemath{\rmatI}{\bimI}
\safemath{\rmatJ}{\bimJ}
\safemath{\rmatK}{\bimK}
\safemath{\rmatL}{\bimL}
\safemath{\rmatM}{\bimM}
\safemath{\rmatN}{\bimN}
\safemath{\rmatO}{\bimO}
\safemath{\rmatP}{\bimP}
\safemath{\rmatQ}{\bimQ}
\safemath{\rmatR}{\bimR}
\safemath{\rmatS}{\bimS}
\safemath{\rmatT}{\bimT}
\safemath{\rmatU}{\bimU}
\safemath{\rmatV}{\bimV}
\safemath{\rmatW}{\bimW}
\safemath{\rmatX}{\bimX}
\safemath{\rmatY}{\bimY}
\safemath{\rmatZ}{\bimZ}

\safemath{\rmatDelta}{\bimDelta}
\safemath{\rmatLambda}{\bimLambda}
\safemath{\rmatPhi}{\bimPhi}
\safemath{\rmatSigma}{\bimSigma}
\safemath{\rmatOmega}{\bimOmega}
\safemath{\rmatTheta}{\bimTheta}

\usepackage{amssymb}
\usepackage{amsfonts}
\usepackage{mathrsfs}
\usepackage{xspace}
\usepackage{bm}
\usepackage{fancyref}
\usepackage{textcomp}

\usepackage{multirow}
\usepackage{stmaryrd}

\newenvironment{textbmatrix}{	\setlength{\arraycolsep}{2.5pt}%
	\big[\begin{matrix}}{\end{matrix}\big]%
	\raisebox{0.08ex}{\vphantom{M}}}

\def\be{\begin{equation}}
	\def\ee{\end{equation}}
\def\een{\nonumber \end{equation}}
\def\mat{\begin{bmatrix}}
\def\emat{\end{bmatrix}}
\def\btm{\begin{textbmatrix}}
\def\etm{\end{textbmatrix}}

\def\ba#1\ea{\begin{align}#1\end{align}}
\def\bas#1\eas{\begin{align*}#1\end{align*}}
\def\bs#1\es{\begin{split}#1\end{split}}
\def\bg#1\eg{\begin{gather}#1\end{gather}}
\def\bml#1\eml{\begin{multline}#1\end{multline}}
\def\bi#1\ei{\begin{itemize}#1\end{itemize}}

\newcommand{\lefto}{\mathopen{}\left}

\DeclareMathOperator*{\argmin}{arg\;min}		% arg min
\DeclareMathOperator{\Exop}{\opE}			% expectation operator
\DeclareMathOperator{\Varop}{\opV\!\mathrm{ar}} % variance operator
\newcommand{\Ex}[2]{\ensuremath{\Exop_{#1}\lefto[#2\right]}} 	% expectation
\newcommand{\Var}[1]{\ensuremath{\Varop\lefto[#1\right]}} % variance
\safemath{\dirac}{\delta}					% Dirac delta
\safemath{\krond}{\dirac}					% Kronecker delta
\safemath{\upto}{\uparrow}
\safemath{\downto}{\downarrow}
\safemath{\iu}{j}							% imaginary unit
\safemath{\ev}{\lambda}						% eigenvalue
\safemath{\hilseqspace}{l^{2}}				% Hilbert sequence space
\newcommand{\banachfunspace}[1]{\setL^{#1}}	% Banach function space
\safemath{\hilfunspace}{\banachfunspace{2}}	% Hilbert function space
\safemath{\SNR}{\textit{SNR}} 				% signal to noise ratio
\safemath{\PAR}{\textit{PAR}} 				% signal to noise ratio
\safemath{\No}{N_0}							% noise spectral density
\safemath{\Es}{E_s}							% energy per symbol
\safemath{\Eb}{E_b}							% energy per bit
\safemath{\EbNo}{\frac{\Eb}{\No}}
\safemath{\EsNo}{\frac{\Es}{\No}}

\DeclareMathOperator{\CHop}{\ensuremath{\opH}} % channel operator
\safemath{\tvir}{\rndh_{\CHop}}				% time-varying impulse response
\safemath{\tvtf}{\rndl_{\CHop}}				% 	-''- transfer function
\safemath{\spf}{\rnds_{\CHop}}				% spreading function
\safemath{\bff}{H_{\CHop}}					% bi-freuqency function

\safemath{\ircf}{r_{h}}						% impulse response correlation fn.
\safemath{\tftvcf}{r_{s}}					% scattering function
\safemath{\tfcf}{r_{l}}						% time-frequency correlation fn.
\safemath{\bfcf}{r_{H}}						% bi-frequency correlation fn.

\safemath{\tcorr}{c_h}						% time-correlation function
\safemath{\scf}{c_{s}}						% spreading function
\safemath{\tfcorr}{c_{l}}					% transfer-function correlation
\safemath{\fcorr}{c_{H}}						% frequency-correlation function

\safemath{\mi}{I}							% mutual information
\safemath{\capacity}{C}						% capacity

\safemath{\normal}{\mathcal{N}}			% normal distribution
\safemath{\jpg}{\mathcal{CN}}			% jointly proper Gaussian
\safemath{\mchain}{\leftrightarrow}		% Markov chain
\safemath{\dB}{\,\mathrm{dB}}
\safemath{\dBm}{\,\mathrm{dBm}}
\safemath{\Hz}{\,\mathrm{Hz}}
\safemath{\kHz}{\,\mathrm{kHz}}
\safemath{\MHz}{\,\mathrm{MHz}}
\safemath{\GHz}{\,\mathrm{GHz}}
\safemath{\s}{\,\mathrm{s}}
\safemath{\ms}{\,\mathrm{ms}}
\safemath{\mus}{\,\mathrm{\text{\textmu}s}}
\safemath{\ns}{\,\mathrm{ns}}
\safemath{\ps}{\,\mathrm{ps}}
\safemath{\meter}{\,\mathrm{m}}
\safemath{\mm}{\,\mathrm{mm}}
\safemath{\cm}{\,\mathrm{cm}}
\safemath{\m}{\,\mathrm{m}}
\safemath{\W}{\,\mathrm{W}}
\safemath{\mW}{\, \mathrm{mW}}
\safemath{\J}{\,\mathrm{J}}
\safemath{\K}{\,\mathrm{K}}
\safemath{\bit}{\,\mathrm{bit}}
\safemath{\nat}{\,\mathrm{nat}}

\safemath{\define}{\triangleq}			% definition

\safemath{\equivalent}{\sim}
\safemath{\distas}{\sim}					% distributed according to
\safemath{\sdiff}{\Delta}				% symmetric set difference

\safemath{\reals}{\mathbb{R}}
\safemath{\positivereals}{\reals_{+}}
\safemath{\integers}{\mathbb{Z}}
\safemath{\posint}{\integers_{+}}
\safemath{\naturals}{\mathbb{N}}
\safemath{\posnaturals}{\naturals_{+}}
\safemath{\complexset}{\mathbb{C}}
\safemath{\rationals}{\mathbb{Q}}

\newcommand*{\fancyrefapplabelprefix}{app}		% Appendix
\newcommand*{\fancyrefthmlabelprefix}{thm}		% Theorem
\newcommand*{\fancyreflemlabelprefix}{lem}		% Lemma
\newcommand*{\fancyrefcorlabelprefix}{cor}		% Corollary
\newcommand*{\fancyrefdeflabelprefix}{def}		% Definition
\newcommand*{\fancyrefproplabelprefix}{prop}		% Proposition
\newcommand*{\fancyrefexmpllabelprefix}{exmpl}
\newcommand*{\fancyrefalglabelprefix}{alg}		% Algorithm
\newcommand*{\fancyreftbllabelprefix}{tbl}		% Algorithm

\frefformat{vario}{\fancyrefseclabelprefix}{Section~#1}
\frefformat{vario}{\fancyrefthmlabelprefix}{Theorem~#1}
\frefformat{vario}{\fancyreftbllabelprefix}{Table~#1}
\frefformat{vario}{\fancyreflemlabelprefix}{Lemma~#1}
\frefformat{vario}{\fancyrefcorlabelprefix}{Corollary~#1}
\frefformat{vario}{\fancyrefdeflabelprefix}{Definition~#1}
\frefformat{vario}{\fancyreffiglabelprefix}{Figure~#1}
\frefformat{vario}{\fancyrefapplabelprefix}{Appendix~#1}
\frefformat{vario}{\fancyrefeqlabelprefix}{(#1)}
\frefformat{vario}{\fancyrefproplabelprefix}{Proposition~#1}
\frefformat{vario}{\fancyrefexmpllabelprefix}{Example~#1}
\frefformat{vario}{\fancyrefalglabelprefix}{Algorithm~#1}

\usepackage{color}

\usepackage{enumitem}
\setlist[itemize]{leftmargin=*, itemsep=0.3em, topsep=0.3em} 

\allowdisplaybreaks % THIS ALLOWS EQUATIONS TO BREAK THE PAGE

\safemath{\Tran}{\textnormal{T}}
\safemath{\Herm}{\textnormal{H}}

\newtheorem{remark}{Remark}

\begin{document}

\title
{Beamspace Equalization for mmWave Massive MIMO: Algorithms and VLSI Implementations}
\author{Seyed Hadi Mirfarshbafan and Christoph Studer%
\thanks{This paper completes and extends the results from the authors' earlier work in \cite{mirfarshbafan21b,SeyedHadi20b}. Concretely, we propose an improved version of SPADE from~\cite{mirfarshbafan21b}, and we develop new VLSI architectures and implementations for the improved algorithm as well as the beamspace data-detection algorithms from \cite{SeyedHadi20b}.}
\thanks{This work was supported in part by the Swiss National Science Foundation (SNSF) grant 200021\_207314, by CHIST-ERA grant for the project CHASER (CHIST-ERA-22-WAI-01) through the SNSF grant 20CH21\_218704, and by the European Commission within the context of the project 6G-REFERENCE (6G Hardware Enablers for Cell Free Coherent Communications and Sensing), funded under EU Horizon Europe Grant Agreement 101139155.}
		\thanks{Seyed Hadi Mirfarshbafan and Christoph Studer are with the Department of Information Technology and Electrical Engineering, ETH Zurich, Switzerland. (email: mirfarshbafan@iis.ee.ethz.ch and studer@ethz.ch)}		

}

\maketitle

\begin{abstract}
	
	Massive multiuser multiple-input multiple-output (MIMO) and millimeter-wave (mmWave) communication are key physical layer technologies in future wireless systems. Their deployment, however, is expected to incur excessive baseband processing hardware cost and power consumption. Beamspace processing leverages the channel sparsity at mmWave frequencies to reduce baseband processing complexity. In this paper, we review existing beamspace data detection algorithms and propose new algorithms as well as  corresponding VLSI architectures that reduce data detection power. We present VLSI implementation results for the proposed architectures in a 22\,nm FDSOI process. Our results demonstrate that a fully-parallelized implementation of the proposed complex sparsity-adaptive equalizer (CSPADE) achieves up to 54\% power savings compared to antenna-domain equalization. 
	Furthermore, our fully-parallelized designs achieve the highest reported throughput among existing massive MIMO data detectors, while achieving better energy and area efficiency. 
	We also present a sequential multiply-accumulate (MAC)-based architecture for CSPADE, which enables even higher power savings, i.e., up to 66\%, compared to a MAC-based antenna-domain equalizer. 
\end{abstract}


\section{Introduction}

\IEEEPARstart{M}{illimeter}-wave (mmWave) frequency bands are being deployed in modern wireless communication devices and have already been adopted in the 5th generation (5G) new radio standard \cite{3gpp22}. A major challenge of communication at these frequencies is the high path loss~\cite{rappaport15a}. Massive multiple-input  multiple-output (MIMO)~\cite{larsson14a} is a key physical layer technology in 5G and beyond 5G, which can mitigate the path-loss issue through beamforming with large antenna arrays.

In all-digital basestation (BS) architectures, each BS antenna is equipped with a dedicated radio-frequency (RF) chain. Although such architectures are more expensive and consume more power compared to hybrid analog-digital beamforming architectures~\cite{sohrabi16, heath-jr.15a}, they achieve higher spectral efficiency, provide more flexibility, and simplify radio-frequency (RF) circuitry and baseband processing~\cite{yan2019performance, panagiotis20}.
Nonetheless, baseband processing in all-digital wideband mmWave massive MU-MIMO systems is challenging due to (i) large sampling rates to support high bandwidth communication and (ii) the high-dimensional signals acquired in massive MIMO. 

An emerging approach to reduce power consumption of baseband processing in all-digital mmWave massive MU-MIMO is to exploit beamspace sparsity \cite{schniter14a, mirfarshbafan19a} of mmWave channels to reduce baseband complexity~\cite{Abdelghany19, Mahdavi20, SeyedHadi20b, mirfarshbafan21b, goluntas21}. This approach, commonly referred to as \emph{beamspace processing},  is based on the fact that at higher carrier frequencies, electromagnetic waves' propagation becomes predominantly directional, and fewer reflected paths arrive at the receiver antennas compared to communication sub-6-GHz frequencies~\cite{rappaport15a}. Therefore, taking a discrete Fourier transform (DFT) across the BS antenna array for each baseband received vector yields an approximately sparse beamspace vector, which can be leveraged to reduce baseband processing complexity and power consumption. 

\subsection{Related Work} \label{sec:PreviousWork}
Earlier works that exploited mmWave beamspace sparsity to reduce complexity focused mainly on single-user MIMO systems with hybrid analog-digital front ends \cite{CAPMIMO, Song13Beamspace}. The case of multi-user hybrid MIMO systems was studied in \cite{SayeedGLOBECOM, GaoNearOptimalBeamspace}.
Beamspace processing in all-digital architectures has been studied only recently. 
In what follows, we focus on beamspace processing for uplink data detection---similar methods can be developed for downlink precoding; see, e.g.,~\cite{goluntas21}.

One of the earliest all-digital massive MU-MIMO beamspace data detectors was proposed in \cite{Abdelghany19}, which is called the ``local linear minimum mean-square error (LMMSE)'' algorithm. This algorithm identifies a contiguous window of $W$ discrete beams for each UE, so that the resulting linear filter minimizes the mean squared error (MSE) of the estimated symbol for each UE independently. 
In contrast to local LMMSE, which selects a different subset of beams for each UE, two alternative beam-selection methods were proposed in \cite{MahdaviGlobalSIP}, which select a subset of beams to be used for data detection across all UEs. 
The first algorithm, referred to as the ``largest columns (LC)'' algorithm (consistent with the terminology in~\cite{SeyedHadi20b}), selects the beams with the largest power averaged over all UEs, while the second algorithm selects the beams with the highest power variance over UEs. A related algorithm was proposed in \cite{Mahdavi20}, where the beams with the largest number of \emph{strong UEs}, i.e., UEs whose power in that beam exceeds the average power of that beam across all UEs, are selected. In what follows, we refer to this algorithm as ``strongest beams (SB)'' algorithm. 

In \cite{SeyedHadi20b}, the beam-selection problem was formulated as an optimization problem, and two approximate algorithms based on orthogonal matching pursuit (OMP)~\cite{OMP} were proposed, namely the ``entry-wise OMP (EOMP)'' and ``column-wise OMP (COMP)''. Furthermore, reference~\cite{SeyedHadi20b} compared those OMP-based algorithms with local LMMSE as well as the LC and SB algorithms, and it was demonstrated that EOMP achieves the best performance-complexity trade-off. 

In \cite{mirfarshbafan21b}, a different approach was proposed to exploit beamspace sparsity to reduce the computational complexity.
In this scheme, referred to as sparsity-adaptive equalization (SPADE), instead of designing a new data detection algorithm, the beamspace LMMSE filter matrix is computed first. 
Then, the effective number of multiplications involved for equalization is reduced by dynamically skipping insignificant multiplications. This scheme was shown to offer significant power savings, while incurring only a small error-rate performance loss. SPADE and the OMP-based methods from~\cite{SeyedHadi20b} will be reviewed in \fref{sec:beamspace_algorithms}, where we also propose an improved version of SPADE we call ``complex SPADE (CSPADE).''

In \cite{Yoshida22}, an algorithm that builds upon local LMMSE was proposed, which improves its error-rate performance at the cost of small extra computations based on log-likelihood ratio (LLR) values. 
Both \cite{Abdelghany19} and \cite{Yoshida22} rely on the Schur complement to reduce the complexity of matrix inversion. We note, however, that Schur complement-based matrix inversion requires high-precision divisions and is susceptible to error accumulation, which renders fixed-point hardware implementations difficult. 
Aside from the aforementioned challenges, the data detector in \cite{Abdelghany19} is evaluated only with line-of-sight (LoS) channel vectors. Similarly, the data detector in \cite{Yoshida22} is evaluated using a geometric one-ring channel model with $10$ degrees angular spread for each transmitter, which does not account for realistic scattering which can occur in non-LoS channels. In contrast, we focus on computationally-efficient methods and evaluate our algorithms using realistic LoS and non-LoS channels. 

We emphasize that existing beamspace data detection methods---except for \cite{Mahdavi20} to which we compare our results in \fref{sec:BER} and \fref{sec:impl_comp}---do not present very-large scale integration (VLSI) implementation results and rely on multiplication counts as a measure of complexity. Such results do not accurately characterize the underlying performance-complexity trade-offs.
In contrast, we will focus on hardware-friendly beamspace data detection algorithms and evaluate the underlying performance-power trade-offs using post-layout VLSI implementation results.

\subsection{Contributions}

The main goal of this paper is to compare the power-performance trade-off of antenna-domain and beamspace equalizers based on post-layout VLSI implementation results. Our contributions are summarized as: 
\begin{itemize}
	\item We propose CSPADE, an improved variant of SPADE. Our VLSI implementation results demonstrate that CSPADE enables more power savings than SPADE.
	\item  We evaluate the performance-complexity trade-off of the state-of-the-art and the proposed beamspace data detection algorithms using Monte-Carlo simulations with realistic LoS and non-LoS channels.

	\item We develop parallel adder tree-based as well as sequential multiply-accumulate-based VLSI architectures for antenna-domain and beamspace equalization. 
	We design mute-capable multipliers which enable power savings for beamspace equalization.
	
	\item We optimize the bitwidths of key signals in each design to minimize their area and power. We demonstrate that beamspace signals exhibit a larger dynamic range (as a consequence of sparsity) compared to antenna-domain signals, and hence require larger bitwidths, which results in higher area and power of beamspace equalizers. 
	\item 
	We present VLSI implementation results and power-performance trade-offs for beamspace and antenna-domain data detectors. We demonstrate that CSPADE achieves the highest-in-class throughput, while achieving similar or better energy-efficiency compared to existing data detector designs.
\end{itemize}

\subsection{Notation} \label{sec:notation}

We use boldface lowercase and uppercase letters to represent column vectors and matrices, respectively. For a matrix $\bA$, the transpose and Hermitian transpose are denoted by $\bA^\Tran$ and $\bA^\Herm$, respectively.
The $m$th column of a matrix $\bA$ is $\bma_m = [\bA]_m$, the transpose of the $n$th row is~$\bma^\text{r}_n = [\bA^\Tran]_{n,:}$, and the entry on the $m$th row and $n$th column  is $A_{m,n} = [\bA]_{m,n}$.
The $N\times N$ identity matrix is~$\bI_N$. The unitary $N\times N$ discrete Fourier transform matrix is denoted by $\bF$ and satisfies $\bF\bF^H=\bI_N$.
For a vector $\bma$, the $k$th entry is denoted by $a_k = [\bma]_k$, and the real and imaginary parts  are denoted by $\Re(\bma)=\bma^\mathcal{R}$ and $\Im(\bma)=\bma^\mathcal{I}$, respectively.
The $\ell_2$-norm of $\bma$ is denoted by~$\|\veca\|_2$, and the $\ell_\infty$-norm and $\ell_{\widetilde\infty}$-norm~\cite{seethaler10a} are defined as $\|\bma\|_\infty \define \max_{k} |a_k|$ and $\|\bma\|_{\widetilde\infty} \define \max\{\|\bma^{\mathcal{R}}\|_\infty,\|\bma^{\mathcal{I}}\|_\infty\}$, respectively.
We use a bar (e.g., $\bar{\bma}$) over quantities in the antenna domain to distinguish them from beamspace vectors or matrices.

\subsection{Paper Outline}
The rest of this paper is organized as follows. \fref{sec:SysModel} introduces the system model and reviews existing beamspace detection algorithms. \fref{sec:sim}  investigates performance-complexity trade-offs via simulations and presents a fixed-point precision analysis for  antenna-domain and beamspace signals. \fref{sec:vlsi} presents VLSI architectures for the proposed beamspace algorithms and  \fref{sec:impl_res} shows corresponding post-layout VLSI implementation results. \fref{sec:conclusion} concludes.

\section{Preliminaries} \label{sec:SysModel}

\begin{figure}[tp]
	\centering
	\includegraphics[width=0.9\columnwidth]{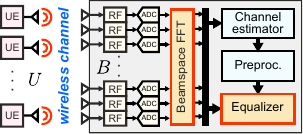}
	%\vspace{-0.2cm}
	\caption{Overview of the considered all-digital massive MU-MIMO uplink system. The spatial beamspace FFT exists only in beamspace  architectures.}
	\label{fig:system_overview}
\end{figure}

We now introduce the antenna-domain and beamspace system models, and we review the beamspace data detection algorithms from~\cite{SeyedHadi20b,mirfarshbafan21b}.
Throughout the paper, we consider the uplink of an all-digital massive MU-MIMO system, in which $U$ single-antenna user equipments (UEs) transmit data to a basestation (BS) with a uniform linear array (ULA) of~$B$ antennas, as depicted in \fref{fig:system_overview}. The UEs transmit their data simultaneously and at the same frequency band.  In an all-digital architecture, each antenna is connected to a dedicated radio-frequency (RF) chain and a pair of analog to digital converters (ADCs) (for simplicity, only one ADC is shown per antenna in \fref{fig:system_overview}). 
The main drawback of all-digital massive MIMO BSs is the high hardware complexity and cost due to the large number of RF chains and ADCs, which can be partially alleviated by using lower-resolution ADCs \cite{risi2014massive,studer16a, mirfarshbafan20}. Furthermore, reducing the ADC resolution can help to reduce the complexity and power consumption of the baseband processor, due to the lower resolution of incoming signals. Therefore, in this paper we optimize the ADC resolution, as well as other relevant baseband signals in order to minimize the implementation complexity and power consumption, while minimizing the resulting performance degradation. These optimizations are detailed in \fref{sec:fixp}.

In what follows, we assume frequency-flat and block-fading wireless channels, i.e., the channel stays constant during a coherence interaval of $T$ transmission time slots. Within each coherence interval, the UEs first transmit orthogonal pilots for channel estimation, followed by data transmission.\footnote{With orthogonal frequency division multiplexing (OFDM), the results of this paper can be extended to frequency-selective channels.}

\subsection{Antenna-Domain System Model}

Let $\bms \in\setS^U $ denote the vector of data symbols transmitted by all UEs during one time slot. The data symbols are chosen from a discrete constellation set $\setS$, e.g., 16-QAM, with the power constraint $\Ex{}{|s_u|^2} = E_s$, $u=1,\ldots,U$. The vector of baseband received signals at the input of the ADCs is given by 
\begin{align} \label{eq:AD_model}
	\bar{\bmz} = \bar{\bH}\bms + \bar{\bmn},
\end{align} 
where $\bar{\bH} \in \complexset^{B\times U}$ is the channel matrix for a given coherence interval, and $\bar{\bmn}$ is the noise modeled as independent and identically distributed (i.i.d.) circularly-symmetric complex Gaussian random vector with variance $\No$ per entry. 
\subsection{Analog-to-Digital Converter (ADC) Model} \label{sec:adc_model}

In all-digital BS implementations, each antenna port is connected to a pair of ADCs, which sample the quadrature and in-phase components of the received signal independently. The input to the ADCs of the $b$-th antenna are $\Re(\bar{z}_b)$ and $\Im(\bar{z}_b)$, where $\bar{z}_b = (\bar{\bmh}^r_b)^\Tran \bms + n_b$, with $\bar{\bmh}^r_b$ denoting a column vector which is the transpose of the $b$th row of $\bar{\bH}$. 
In massive MU-MIMO, due to the large number of simultaneously transmitting UEs, we can approximate the distribution of $\bar{z}_b$ (for a fixed channel matrix~$\bar{\bH}$) as Gaussian, using the central limit theorem.

We model the quantization effect of an $m$-bit ADC using the  uniform symmetric quantization function $y = \setQ_m(z, \Delta)$~\cite{sven_thesis},
where $\Delta$ is the quantizer step size.
For Gaussian inputs, it is possible to derive the step size which minimizes the mean squared error (MSE) between the quantizer input $z$ and output~$y$ given by $\text{MSE} = \Ex{z}{(\setQ_m(z, \Delta) - z)^2}$. For zero-mean Gaussian inputs with variance $\sigma^2$, the MSE-optimal step size is $\Delta = \sigma \Delta_1$, where $\Delta_1$ is the MSE-optimal step size for standard Gaussian inputs, which is tabulated in~\cite[Tbl.~3.1]{sven_thesis} for a range of quantizer bits ($m$). 
Adapting the quantizer step size to the variance of its input (i.e., $\Delta = \sigma \Delta_1$) mimics the operation of an automatic gain controller (AGC) employed at the input of ADCs.\footnote{In practice,  ADCs have fixed quantization levels, which corresponds to a quantizer with a fixed step size. An AGC scales the incoming signal to the ADC's dynamic range, rather than scaling the step size of the ADC. Nonetheless, these two approaches are equivalent for our analysis purpose.}
In this paper, we model the vector of received signals at the output of the ADCs as
\begin{align} \label{eq:AD_quant}
	\bar{\bmy} = \setQ_m(\bar{\bmz}, \Delta) = \setQ_m(\bar{\bH}\bms + \bar{\bmn}, \Delta),
\end{align} 
where $\setQ_m(.,\Delta)$ operates element-wise on the input vector and independently for the real and imaginary parts, i.e., $\setQ_m(\bar{z}, \Delta) = \setQ_m(\bar{z}^\mathcal{R}, \Delta) + j\setQ_m(\bar{z}^\mathcal{I}, \Delta)$. For simplicity, we use the same step size $\Delta$ for all entries of the vector and for both the real and imaginary components. The value of the unified step size is given by $\Delta = \max_{b = 1, \ldots B} \{\Delta_1\sqrt{\Var{\bar{z}_b}/2}\}$, where $\Var{\bar{z}_b} = E_s\|\bar{\bmh}_b^{\text{r}}\|_2^2 + \No$, is the variance of the complex received signal at the $b$th BS antenna. This  step-size choice avoids excessive signal clipping  and performs well in practice.

\subsection{Beamspace System Model}
\label{sec:beam_sys}

By assuming mmWave carrier frequencies and that the UEs and scatterers are sufficiently far apart from the BS, we can model the channel between a given UE and the BS according to the well-known plane-wave approximation~\cite{tse05a}:
\begin{align}
	\label{eq:planarwave}
	\bar{\bmh} = \sum_{\ell=0}^{L-1} \alpha_\ell \bar{\bma}(\phi_\ell).
\end{align}
Here, $L$ is the number of propagation paths, $\alpha_l\in\opC$ stands for the channel gain of the $\ell$th propagation path, and
\begin{align} \label{eq:complexsinusoids}
	\bar{\bma}(\phi_\ell) = \big[1, e^{j\phi_\ell},e^{j2\phi_\ell},\dots, e^{j(B-1)\phi_\ell} \big]^\Tran,
\end{align}
where the spatial frequency $\phi_\ell$ is determined by the $\ell$th path's angle-of-arrival to the BS antenna array. At mmWave frequencies, wave propagation is predominantly directional with only a few non-line-of-sight paths. Therefore, for mmWave massive MIMO basestations, $L$ is typically smaller than $B$, which implies that the channel of each UE consists only of a few complex sinusoids. Hence, taking the DFT of $\bar{\bmh}$ in \fref{eq:planarwave} results in an approximately sparse vector (i.e., the magnitudes of most entries are close to zero) \cite{mirfarshbafan19a}. 
By applying the DFT matrix to both sides of \fref{eq:AD_model}, we obtain the following (unquantized) beamspace system model:
\begin{align} \label{eq:BD_model}
	\bmz = \bF\bar{\bmz} = \bF\bar{\bH}\bms + \bF\bar{\bmn}
	= \bH \bms + \bmn.
\end{align} 
Here,  $\bH = \bF\bar{\bH}$ is the beamspace channel matrix and $\bmn = \bF\bar{\bmn}$ is the beamspace-equivalent noise vector, which has the same statistics as $\bar{\bmn}$ since $\bF$ is unitary. 
Each row $b=1, \ldots, B$ of the beamspace channel matrix $\bH$ and received vector $\bmz$ correspond to one spatial beam with main lobe towards a specific angle-of-arrival (AoA); hence the indices~$b$ are referred to as \emph{beam indices}. 
Furthermore, since~$\bF$ is unitary, the unquantized beamspace system model \fref{eq:BD_model} is mathematically equivalent to~\fref{eq:AD_model} and data detection in both domains can give the same result. 
In practice, the beamspace DFT is applied on the quantized received vector, i.e., \emph{after} the ADCs. Thus, the beamspace received vector available for uplink processing is given by
\begin{align} \label{eq:BD_quant}
	\bmy = \bF \setQ_m(\bar{\bH}\bms + \bar{\bmn}, \Delta).
\end{align} 

\subsection{Linear Data Detection}
\label{sec:lindetection}

Most practical massive MIMO receivers resort to linear data detection to minimize computational complexity~\cite{Fang17, prabhu17, Mahdavi20, mirfarshbafan24}. 
Linear data detection methods typically operate in two phases: \emph{preprocessing} and \emph{equalization}. 
During preprocessing, which is performed once per channel coherence interval, an equalization filter matrix~$\bar{\bW}$ is computed from the estimated channel matrix~$\hat{\bar{\bH}}$.
During equalization, which is performed for every received vector ($T$ times per coherence interval), estimates  $\hat{\bms} = \bar{\bW}\bar{\bmy}$ of the transmit vector are generated. 
For antenna-domain detection, we focus on the widely-used antenna-domain LMMSE equalizer, which we refer to as \emph{ALMMSE}. 
The antenna-domain LMMSE equalization filter can be obtained by solving
\begin{align} \label{eq:ALMMSE}
	\bar{\bW} =\argmin_{\tilde{\bW}\in\complexset^{U\times B}} \| \bI_U - \tilde{\bW} \bar{\bH}\|_F^2+\rho \| \tilde{\bW} \|_F^2,
\end{align} 
where $\rho = N_0/E_s$. The solution to \fref{eq:ALMMSE} is given by \mbox{$\bar{\bW} = (\bar{\bH}^\Herm \bar{\bH} + \rho \bI)^{-1}\bar{\bH}^\Herm$}. In practice, one uses the estimated channel matrix $\hat{\bar{\bH}}$ instead of the true~$\bar{\bH}$.

Beamspace equalization resembles the above procedure with the exception that the equalization matrix $\bW$ is computed from the estimated beamspace channel matrix $\hat{\bH}$, and is applied to the beamspace receive vectors~$\bmy$ in order to obtain the transmitted symbol estimates $\hat{\bms} = {\bW}{\bmy}$. 
Beamspace equalization requires each antenna-domain received vector to be converted to beamspace via a DFT. Beamspace LMMSE which we refer to as \emph{BLMMSE} is the solution to the following problem:
\begin{align} \label{eq:BLMMSE}
	\bW =\argmin_{\tilde{\bW}\in\complexset^{U\times B}} \| \bI_U - \tilde{\bW} \bH\|_F^2+\rho \| \tilde{\bW} \|_F^2.
\end{align}

\subsection{Beamspace Sparsity}
\label{sec:sparsity}

\begin{figure}[tp]
	\centering	
	\begin{subfigure}{0.5\columnwidth}
		\includegraphics[width=\columnwidth]{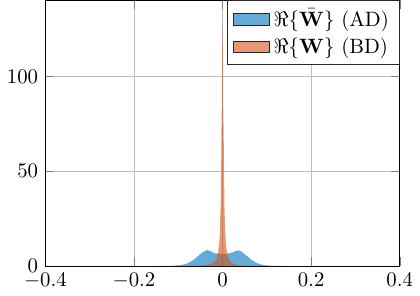}
		\caption{PDF of $\Re\{\bar{\mathbf{W}}\}$ and $\Re\{\mathbf{W}\}$.}
		\label{fig:PDF_W64_LOS}
	\end{subfigure}
	\hfill
	\begin{subfigure}{0.48\columnwidth}
		\includegraphics[width=\columnwidth]{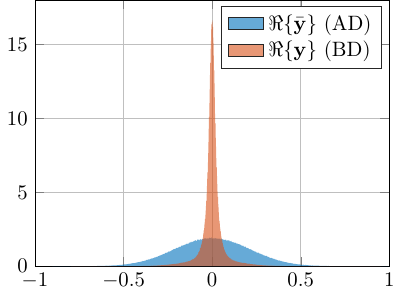}
		\caption{PDF of $\Re\{\bar{\mathbf{y}}\}$ and $\Re\{\mathbf{y}\}$.}
		\label{fig:PDF_Y64_LOS}
	\end{subfigure}
	\caption{Empirical PDF of real part of entries of (a) $\bar{\bW}$/${\bW}$ and (b) $\bar{\bmy}$/${\bmy}$ using LoS channels generated by QuaDRiGa. BD stands for beamspace domain and AD for antenna domain.}
	\label{fig:PDF}
	\vspace{-0.2cm}
\end{figure}

As mentioned in \fref{sec:beam_sys}, mmWave massive MIMO channels are sparse in beamspace.
Due to channel hardening in massive MIMO systems, the regularized Gram matrix $\bA = \bH^\Herm \bH + \rho \bI$ tends to be diagonally dominant; thus, the beamspace LMMSE matrix $\bW = \bA^{-1}\bH^\Herm$ also exhibits sparsity.
This sparsity property is illustrated in \fref{fig:PDF_W64_LOS}, which shows the empirical probability density function (PDF) of the real part of entries of $\bW$ and $\bar{\bW}$. We observe that the PDF of the real part of the entries of the beamspace domain (BD) ${\bW}$ are more concentrated around zero than the entries of the antenna domain (AD) $\bar{\bW}$.
We emphasize that, to be able to compare the PDFs of entries of $\Re\{\bar{\bW}\}$ and $\Re\{\bW\}$, we scaled all instances of $\bar{\bW}$ with a scalar $\bar{\alpha}$ and all instances of $\bW$ with another scalar $\alpha$, to unify the range of entries of $\Re\{\bar{\bW}\}$ and $\Re\{\bW\}$. Similarly, as demonstrated in \fref{fig:PDF_Y64_LOS}, the beamspace receive vector $\bmy$, which is a linear combination of columns of $\bH$ plus noise, is sparser than $\bar{\bmy}$.

The key  advantage of performing equalization in beamspace rather than in the antenna domain is that one can exploit beamspace sparsity in (i) the  LMMSE matrix $\bW$ and (ii) the received vector $\bmy$ to reduce computational complexity and achieve more efficient hardware implementations. 

\subsection{Beamspace Equalization Algorithms} \label{sec:beamspace_algorithms}
For linear data detection in beamspace, there exist two general approaches to exploit sparsity with the goal of reducing computational complexity: (i) design a preprocessing algorithm that produces a strictly sparse equalization matrix with a fixed number of zero elements \cite{Abdelghany19, Mahdavi20, SeyedHadi20b, Yoshida22}; or (ii) use conventional linear data detectors, such as ZF or LMMSE, and exploit the fact that the equalization matrix obtained by such linear detectors in beamspace is approximately sparse and reduce the effective number of multiplications, e.g., using SPADE~\cite{mirfarshbafan21b}.
The beamspace data detection algorithms of the first approach that produce strictly sparse equalization matrices, are governed by a \emph{density coefficient} $\delta \in[0,1]$, which specifies the fraction of nonzero elements out of the total number of elements in the resulting equalization matrix, and affects the error-rate performance. Based on the sparsity structure of the resulting equalization matrix $\bW \in \complexset^{U \times B}$, these algorithms are categorized into \textit{column-wise} and \textit{entry-wise} methods. 
The resulting sparse equalization matrix $\bW$ has either \textit{column-wise} sparsity which means only $\delta B$ out of $B$ columns can be nonzero and the rest must be all-zero columns, or it has \textit{entry-wise} sparsity, meaning that each row of $\bW$ has $\delta B$ nonzero elements, but the support set of each row of $\bW$ can be different. 

In \cite{SeyedHadi20b}, several state-of-the-art beamspace data detection algorithms were compared and it was shown that COMP and EOMP achieve a better performance-complexity trade-off compared to competing algorithms, given sufficiently long channel coherence times. 
We now further explore this performance-complexity trade-off based on actual VLSI implementations and compare them with traditional antenna-domain equalization methods. We now briefly summarize linear beamspace data detectors that achieve competitive performance-complexity trade-offs, i.e., EOMP and COMP, as well as SPADE from~\cite{mirfarshbafan21b}. Finally, we propose a novel variant of SPADE, called CSPADE.
\begin{remark} \label{rmk:1}
	Additional post processing methods for the output of the local LMMSE method from~\cite{Abdelghany19}, were proposed in \cite{Abdelghany20,Yoshida22}, with the aim of improving error-rate performance. Similarly, the beamspace data detector proposed in \cite{Mahdavi20}, consists of a linear beamspace detector with a beam-selection method which we refer to as ``strongest beams (SB),'' followed by a nonlinear post-processing stage to improve its performance. Such nonlinear post-processing extensions can also be applied to EOMP, COMP, and SPADE/CSPADE. In this paper, however, we solely focus  on the implementation of linear data detection in antenna domain and beamspace with the goal of identifying conditions in which beamspace equalization offers power savings when compared to the antenna-domain equalization. 
\end{remark}

\subsubsection{Columnwise Orthogonal Matching Pursuit (COMP) \cite{SeyedHadi20b}}
This algorithm takes the density coefficient~$\delta$ as input and tries to find a solution to \fref{eq:BLMMSE}, such that only $K = \delta B$ columns of~$\bW$ have nonzero elements. The optimal solution involves a search over ${B \choose K}$ possible support sets, which is prohibitive. COMP finds an approximate solution $\hat{\bW}$ in $K$ OMP iterations. We refer to \cite{SeyedHadi20b} for the algorithm details.

\subsubsection{Entrywise Orthogonal Matching Pursuit (EOMP) \cite{SeyedHadi20b}} \label{sec:EOMP}
EOMP constructs a sparse equalization matrix $\hat{\bW}$, each row of which has only $\delta B$ nonzero entries. For each row (corresponding to a UE) $u \in \{1,2,\ldots, U\}$, EOMP performs $K = \delta B$ OMP iterations, each of which computes one nonzero entry of that row. We refer to \cite{SeyedHadi20b} for the algorithm details.

\subsubsection{Sparsity-adaptive equalization (SPADE) \cite{mirfarshbafan21b}} \label{sec:spade} 
In contrast to EOMP and COMP, SPADE does not involve a new preprocessing algorithm to compute a strictly sparse equalization matrix.
Instead, SPADE exploits the inherent approximate sparsity of the beamspace LMMSE matrix $\bW = (\bH^\Herm \bH + \rho \bI)^{-1}\bH^\Herm$ to reduce the computational complexity of equalization~$\hat{\bms} = \bW \bmy$. 

Consider the dot product  $\hat{s}_u = (\bmw_u^\text{r})^\Tran \bmy = \sum_{b=1}^B W_{u,b} y_b$. Each such complex-valued inner product can be decomposed into four real-valued inner products. Consider one of these real-valued inner products, e.g., $ \Re\{\bmw_u^\text{r}\}^\Tran \Re\{\bmy\}$. 
Due to the sparsity of $\Re\{\bmw_u^\text{r}\}$ and $\Re\{\bmy\}$, this inner product involves many partial products with potentially very small magnitudes. 
SPADE's idea is to dynamically skip such insignificant multiplications to reduce dynamic power, without noticeably altering the result of the dot product. SPADE uses two thresholds $\tau_y$ and $\tau_w$ to skip real-valued multiplications for which the absolute value of both operands are below the respective thresholds. These thresholds trade power savings with accuracy and are determined offline based on Monte--Carlo simulations to minimize the approximation error while maximizing the saving in effectively carried out multiplications.

\subsubsection{Complex SPADE (CSPADE)} \label{sec:cspade} 

We now propose an improved variant of  SPADE~\cite{mirfarshbafan21b}, which we call CSPADE. 
Instead of operating on the real-valued decomposition of complex-valued inputs, CSPADE operates directly on the complex-valued inputs, and hence, skips entire complex-valued multiplications rather than individual real-valued multiplications. 
As for SPADE, CSPADE takes two thresholds $\tau_y$ and $\tau_w$ and skips the partial products in $(\bmw_u^\text{r})^\Tran \bmy = \sum_{b=1}^B W_{u,b} y_b$, when the magnitude of both operands are below their respective thresholds. 
A straightforward computation of the  magnitude $\| x\| = \sqrt{(x^\mathcal{R})^2 + (x^\mathcal{I})^2}$ of a complex input $x = x^\mathcal{R} + jx^\mathcal{I}$, would require two multiplications, an addition and a square root operation. Instead, in CSPADE we use the $\ell_{\widetilde\infty}$-norm as a proxy for the magnitude of the complex inputs, which significantly simplifies hardware implementation. A comparison of the $\ell_{\widetilde\infty}$-norm of the input with $\tau$ amounts to simply comparing the absolute value of the real and imaginary parts of the input with $\tau$ and then AND-ing the result. Our VLSI implementation results shown in \fref{sec:pwr_brk} demonstrate that CSPADE enables more power savings than SPADE.

\subsection{Multiplier Activity Rate} \label{sec:activity_rate} 

The density coefficient $\delta$ in EOMP and COMP quantifies the complexity of the equalization operation $\hat{\bms}=\bW \bmy$. Since in these algorithms, $\bW$ has always $\delta BU$ nonzero elements, the equalization operation consists of at most $4 \delta BU$ real-valued multiplications.\footnote{Throughout the paper, we tacitly assume four real-valued multiplications per complex-valued multiplication. Utilizing Gauss' trick~\cite[p.~647]{knuth1981art2} with three real-valued multiplications would change our complexity counts.
}
With SPADE and CSPADE, however,~$\bW$ does not have strict sparsity and the effective number of multiplications in the equalization depends on the specific instances of $\bW$ and $\bmy$, as well as the threshold pair. Therefore, in order to quantify the SPADE and CSPADE equalization complexity, we define the \emph{multiplier activity rate} (for a given threshold pair) as the average number of executed real-valued multiplications divided by $4UB$. We denote the multiplier activity rate by $\alpha$. Note that multiplier activity rate is equivalent to the density coefficient $\delta$ for strictly sparse equalization methods, as they both quantify the ratio between the effective number of multiplications and the total number of multiplications required to perform linear equalization.

\section{Simulation results} \label{sec:sim}

We now present simulation results for the beamspace equalization algorithms EOMP, COMP, SPADE, and CSPADE, and for the traditional ALMMSE data detector. We assess the performance of these algorithms using simulated bit error-rate (BER) and their computational complexity using  the density coefficient $\delta$ and multiplier activity rate $\alpha$.

We simulate uplink data detection in a mmWave massive MU-MIMO BS equipped with a $B$-antenna $\lambda/2$-spaced ULA, in which $U$ single-antenna UEs transmit $16$-QAM symbols to the BS concurrently and in the same frequency band. The channel matrices are generated by QuadRiGa mmMAGIC UMi model \cite{QuaDRiGa_tech_rpt} at a carrier frequency of $60$\,GHz for both LoS and non-LoS scenarios. The UE positions are distributed randomly in a $120^\circ$ sector within $10\,\text{m}$ to $110\,\text{m}$ from the BS array, with a minimum of $1^\circ$ angular separation. We employ BS-side power control so that the variation in receive power of UEs is within the range of~$\pm3$\,dB.

We consider a block-fading scenario, in which the channel stays constant over a block of $T$ transmissions and changes independently across different blocks. The BS first obtains an estimate of the channel during the channel training phase using least-squares (LS) estimation followed by  BEACHES~\cite{mirfarshbafan19a} to denoise the channel estimates. Then, the BS computes the beamspace or antenna-domain equalization matrix from the estimated channel matrix to perform equalization. We model the impact of ADCs during both the channel training and data detection phases, using the ADC model in~\fref{sec:adc_model}.
Our simulations for a wide range of system parameters show that $6$-bit ADCs with uniform symmetric quantization deliver sufficient accuracy (close to the performance with unquantized inputs).
For more hardware-aware simulation results, the beamspace transform FFT is also emulated using the fixed- point implementation of the SMUL-FFT from~\cite{mirfarshbafan21}.

\subsection{Fixed-Point Parameters}
\label{sec:fixp}
We now discuss the fixed-point parameters of the key signals in an equalizer implementation, which are used in both the simulation results presented in this section as well as the VLSI implementation results shown in \fref{sec:impl_res}.
We note that, when we refer to the bitwidth of a complex-valued signal, we mean the individual bitwidth of the real and imaginary part.

A $Q$-bit ADC with uniform symmetric quantization has $2^Q$ output levels which are symmetric around zero. To represent these $2^Q$ levels as a 2's complement fixed-point number, we actually need $Q+1$ bits, since the range of possible values which can be represented in 2's complement is not symmetric around zero. 
Therefore, we need $W=7$ bits for the two's complement fixed-point entries of $\bar{\bmy}$, which come from the $6$-bit ADCs.
For the entries of the antenna-domain LMMSE equalization matrix $\bar{\bW}$, we found that $11$ bits with $10$ fractional bits deliver near-floating-point performance.

As demonstrated in \fref{sec:sparsity}, beamspace signals are sparser than their antenna-domain counterparts. This beamspace sparsity results in a higher dynamic range, which in turn calls for higher resolution in their fixed-point representation compared to antenna-domain signals, to maintain  the same error-rate performance in both domains. 

Based on extensive Monte--Carlo simulations, the fixed-point parameters for beamspace signals we chose are as follows. For the received vector $\bmy = \bF \bar{\bmy}$, we use $W=9$ bits with $F=1$ fractional bits. For the beamspace equalization matrix in all the beamspace equalizers we use $W=12$ bits with $F=11$ fractional bits. 
For both antenna-domain and beamspace equalization, the result of equalization is the same. Therefore, for $\hat{\bms}$, we  use $W=13$ bits with $F=8$ fractional bits.

\fref{tbl:fixp_params} summarizes the fixed-point parameters for the key signals of the VLSI architectures   described in \fref{sec:vlsi}.\footnote{In a fully optimized design, it is possible to further reduce the resolution of $\bar{\bW}$ and $\bW$ with CS down to $9$\,b and $10$\,b, respectively. If we use $7$-bit ADCs instead of $6$-bit, then we can further reduce the $\bar{\bW}$ and $\bW$ resolution down to $8$\,b and $9$\,b, respectively. Note, however, that such a slight improvement of resolution does not change any of the results from this paper.}

\begin{table}[tp]
	\centering
	{\caption{Fixed-point parameters for equalization.}
		\label{tbl:fixp_params} 
		\renewcommand{\arraystretch}{1.5}
		\begin{minipage}[c]{1\columnwidth}
			\centering
			
			\begin{tabular}{@{}lccccc@{}}
				\toprule
				Signal & $\bar{\bmy}$ & $\bar{\bW}$ & $\bmy$ & $\bW$ & $\hat{\bms}$\\
				\midrule
				Bitwidth ($W$) & $7$ & $11$ & $9$ & $12$ & $13$ \\
				Fractional bits ($F$) & $1$ & $10$ & $1$ & $11$ & $8$\\
				\bottomrule
			\end{tabular}
			
	\end{minipage}}
\end{table}

\subsection{Performance-Complexity Trade-offs} \label{sec:tradeoff}

\begin{figure}[tp]
	\centering
	\begin{subfigure}{0.45\linewidth}
		\includegraphics[width=\linewidth]{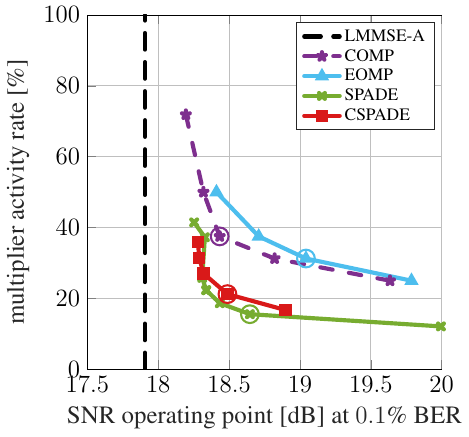}
		\caption{LoS, $B=64$, $U=8$}
		\label{fig:SNRop2:a}
	\end{subfigure}
	\hfill
	\begin{subfigure}{0.45\linewidth}
		\includegraphics[width=\linewidth]{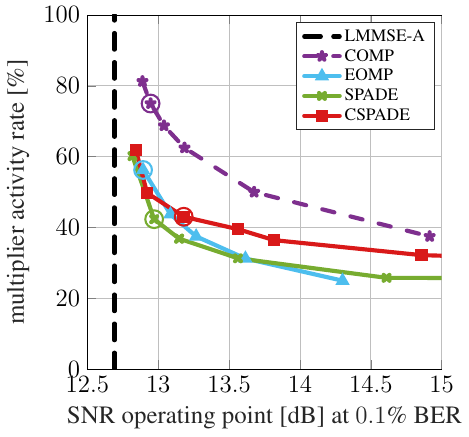}
		\caption{Non-LoS, $B=64$, $U=8$}
		\label{fig:SNRop2:b}
	\end{subfigure}
	
	\begin{subfigure}{0.45\linewidth}
		\includegraphics[width=\linewidth]{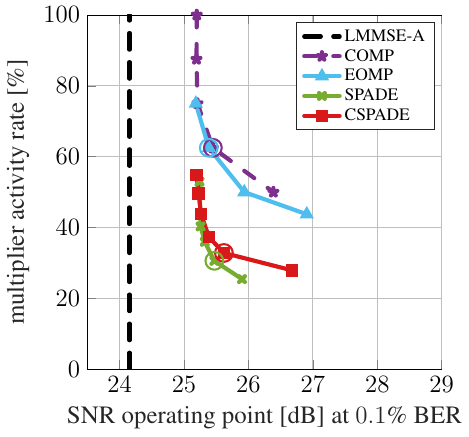}
		\caption{LoS, $B=64$, $U=16$}
		\label{fig:SNRop2:c}
	\end{subfigure}
	\hfill
	\begin{subfigure}{0.45\linewidth}
		\includegraphics[width=\linewidth]{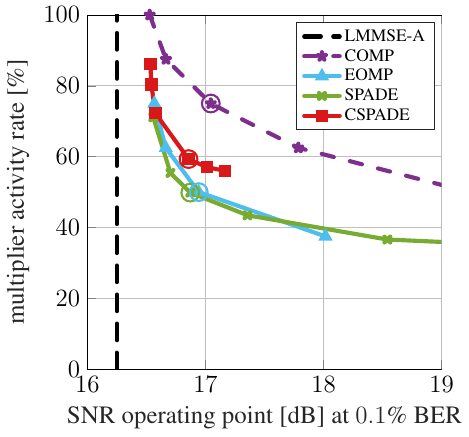}
		\caption{Non-LoS, $B=64$, $U=16$}
		\label{fig:SNRop2:d}
	\end{subfigure}
	
	\caption{Multiplier activity rate (equivalently, the density coefficient for EOMP and COMP) vs.\ SNR operating point at $0.1\%$ uncoded BER. In (a) and (b), the circled data points correspond to the parameters used in the VLSI implementations of each algorithm.}
	\label{fig:SNRop}
	\vspace{-0.2cm}
\end{figure}

We now compare the performance-complexity trade-offs of EOMP, COMP, SPADE, and CSPADE. In what follows, we use the \mbox{ALMMSE} equalizer as our baseline. 

\fref{fig:SNRop} shows the complexity quantified by the multiplier activity rate vs. the SNR operating point (i.e., the minimum SNR to achieve $0.1\%$ uncoded BER), in a $64$-antenna ULA BS receiving $16$-QAM data from $U=\{8,16\}$ UEs in both LoS and non-LoS channels. 
For EOMP and COMP, the multiplier activity rate corresponds to their density coefficient $\delta$.
For SPADE and CSPADE, the performance and complexity are controlled by the employed threshold pair $(\tau_w, \tau_y)$, which gives rise to a specific complexity---quantified by the multiplier activity rate $\alpha$---and performance which is quantified by the SNR operating point.
We performed error-rate simulations for range of threshold pairs for SPADE/CSPADE, and for each pair we computed the SNR operating point and multiplier activity rate. We then chose the best pairs which form the Pareto front plotted in \fref{fig:SNRop}. We emphasize that the results shown in this figure are the fixed-point simulation results with the parameters detailed in \fref{sec:fixp}, and correspond to the implemented hardware of \fref{sec:impl_res}.

\begin{figure*}[tp]
	\centering
	\begin{subfigure}{0.4\linewidth}
		\includegraphics[width=\linewidth]{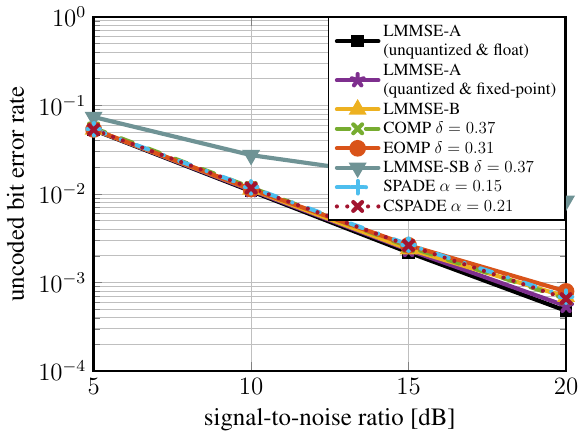}
		\caption{LoS, $B=64$, $U=8$}
		\label{fig:BER:a}
	\end{subfigure}
	\hfill
	\begin{subfigure}{0.4\linewidth}
		\includegraphics[width=\linewidth]{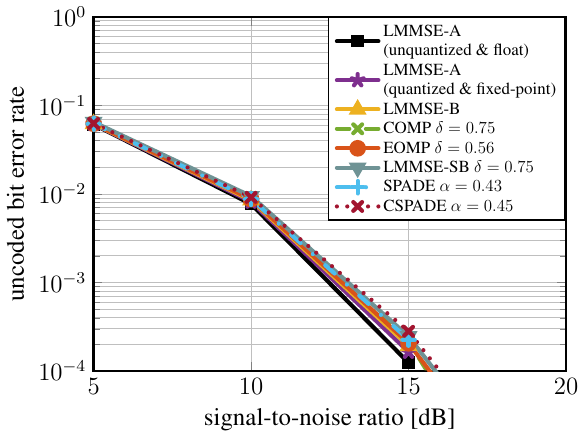}
		\caption{Non-LoS, $B=64$, $U=8$}
		\label{fig:BER:b}
	\end{subfigure}
	
	\begin{subfigure}{0.4\linewidth}
		\includegraphics[width=\linewidth]{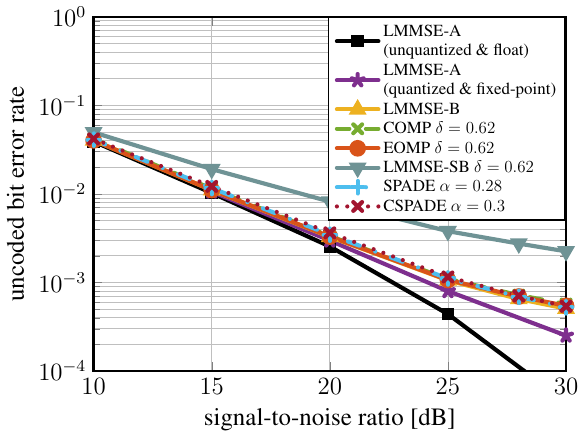}
		\caption{LoS, $B=64$, $U=16$}
		\label{fig:BER:c}
	\end{subfigure}
	\hfill
	\begin{subfigure}{0.4\linewidth}
		\includegraphics[width=\linewidth]{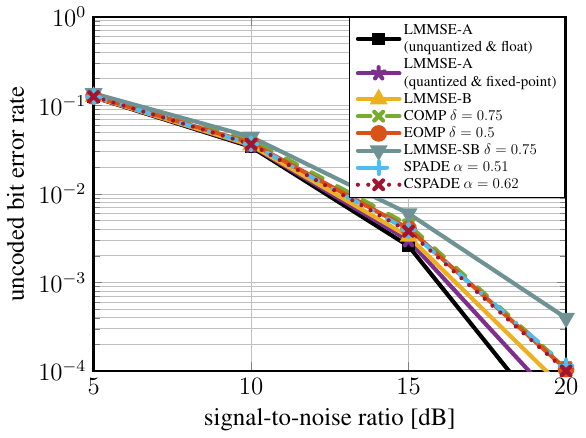}
		\caption{Non-LoS, $B=64$, $U=16$}
		\label{fig:BER:d}
	\end{subfigure}
	
	\caption{Uncoded BER vs SNR for $64\times 8$ and $64 \times 16$ ($B\times U$) systems in LoS and non-LoS channels. The black curve shows the reference BER of floating-point ALMMSE equalization with unquantized inputs, while all other curves show the BER results with inputs quantized with $6$-bit ADCs. Furthermore, all curves except for the ALMMSE  (unquantized \& float) and LMMSE-SB equalizers, show the uncoded BER  considering fixed-point computations.}
	\label{fig:BER}
	\vspace{-0.2cm}
\end{figure*}

As expected, the SNR loss of all beamspace algorithms relative to ALMMSE increases as the multiplier activity rate (or equivalently the density coefficient in EOMP and COMP) goes down. 
Note, however, that there is a residual gap between the fixed-point implementation of the beamspace algorithms and ALMMSE even when the multiplier activity rate approaches $100\%$, which is due to two main factors: (i) beamspace algorithms work with the outputs of the beamspace FFT, whose fixed-point implementation adds an extra layer of quantization error, and (ii) at high SNRs such as in \fref{fig:SNRop2:c}, the beamspace quantities (i.e., entries of $\bW$ and $\bmy$) require even higher resolution in order to close the gap to ALMMSE; therefore the chosen fixed-point parameters described in \fref{sec:fixp} incur a small performance degradation at high SNR.

We emphasize that the impact of fixed-point quantization error on the BER performance is more pronounced at higher SNRs, where the system noise is small. This can be observed in \fref{fig:SNRop2:c}, where higher SNRs are needed to reach the target BER of $10^{-3}$. The SNR gap between the beamspace algorithms and ALMMSE is much smaller at a target BER of $10^{-2}$.

The key takeaway from \fref{fig:SNRop} is that SPADE achieves the best performance-complexity trade-off among the proposed beamspace equalizers, with CSPADE achieving similar trade-offs specially when the SNR gap to ALMMSE is constrained to be small. Furthermore, EOMP achieves a trade-off similar to SPADE in non-LoS channels. We emphasize that the  multiplier activity rate does not fully characterize the implementation complexity. Therefore, in order to truly see which algorithm results in the more power savings in practice, we compare the the simulated power consumption of post-layout  VLSI implementations of these algorithms in \fref{sec:impl_res}. 

\subsection{BER Results} \label{sec:BER}

In order to better illustrate the error-rate performance of the proposed beamspace equalizers, we now also present uncoded BER simulation results.  
\fref{fig:BER} shows the uncoded BER for a system with a $64$-antenna ULA BS serving $8$ or $16$ single-antenna UEs in LoS and non-LoS channel conditions. The baseline is the floating point ALMMSE with unquantized received signals, i.e., no ADC quantization effects. All other curves show BER results for which inputs during both channel estimation and data detection come from $6$-bit ADCs. Several remarks regarding the results are as follows:
\begin{itemize}
	\item \emph{LMMSE-SB}: We include the BER performance of  LMMSE-SB  \cite{Mahdavi20}, which is the only beamspace data detector in the literature for which a VLSI implementation exists. 
	As noted in Remark \ref{rmk:1}, a nonlinear post-processing stage was proposed in \cite{Mahdavi20} to improve its error-rate performance---we do not include such post-processing in our simulations, since we are comparing the performance of linear beamspace equalization algorithms and such nonlinear post-processing could be applied to the output of any data detector. 
	\item \emph{6-bit ADCs:} As we observe, ALMMSE with $6$-bit quantized inputs and fixed-point equalization achieves nearly the same performance as that of ALMMSE with unquantized inputs and floating point equalization up to around $20$\,dB SNR. At higher SNRs, the quantization error becomes dominant as the additive noise variance becomes small.
	\item \emph{BLMMSE:} The beamspace LMMSE equalizer exhibits an SNR gap to the quantized ALMMSE at high SNRs, due to the additional quantization noise resulting from the fixed-point beamspace FFT. Additionally, as discussed in \fref{sec:fixp}, the beamspace quantities require larger bitwidths compared to the corresponding antenna-domain quantities. In our simulations, we optimized the fixed-point parameters to achieve near floating-point performance up to practical SNR values, i.e., SNRs where the uncoded BER goes down to $10^{-3}$. At higher SNRs, we would need an even higher resolution for beamspace signals to close the SNR gap.
	\item \emph{Sparsity parameters:} For EOMP, COMP, CSPADE, and SPADE, we show the BER performance with the smallest density parameters $\delta$ or multiplier activity rate $\alpha$ that achieves good performance, marked by a circle in \fref{fig:SNRop}. For LMMSE-SB, we chose the density coefficient to be the same as for COMP, as both algorithms generate equalization matrices with column-sparsity in~$\bW$. Hence, the equalization implementation is similar for both algorithms. As a consequence, to compare the performance of LMMSE-SB and COMP for the same equalization complexity, we fix the $\delta$ to be equal for both algorithms.
	
\end{itemize}

As one can observe in \fref{fig:BER}, the proposed beamspace equalization algorithms (i.e., EOMP, COMP, CSPADE, and SPADE) all achieve nearly the same BER as BLMMSE; i.e., with the chosen sparsity parameters, these algorithms incur virtually no additional performance degradation due to sparse equalization. In particular, CSPADE and SPADE are able to reduce the multiplier activity rate of equalization down to $21\%$ and $15\%$, respectively, with nearly no performance degradation. Furthermore, we observe that LMMSE-SB suffers a significant performance degradation with the selected density coefficient.

\section{VLSI architectures} \label{sec:vlsi}

In \fref{sec:tradeoff}, we have observed that SPADE and CSPADE offer the best performance complexity trade-off among existing beamspace equalizers.
Furthermore, as shown in \cite{SeyedHadi20b}, EOMP and COMP entail considerably higher preprocessing complexity compared to LMMSE, while as discussed in \fref{sec:spade} and \fref{sec:cspade}, SPADE and CSPADE do not require any additional preprocessing operation compared to LMMSE equalization. 
Hence, SPADE and CSPADE have the additional advantage of requiring simpler preprocessing.

Even when focusing solely on the equalization step, it is not clear from the performance-complexity analysis of \fref{sec:tradeoff}, which algorithm enables greater power savings in hardware implementation, despite the fact that the density coefficient and multiplier activity rate are closely related to the actual hardware complexity of the equalization operation, and, hence, are a good proxy for the potential power savings in hardware. This is because the power saving mechanisms—based on reducing the number of effective multiplications—differ across these algorithms, as will be detailed in this section.

We now present VLSI architectures for the equalization stage, i.e., for computing $\hat{\bms} = \bar{\bW} \bar{\bmy}$ for ALMMSE  and $\hat{\bms} = \bW \bmy$ for the beamspace equalizers: EOMP, SPADE, and CSPADE. For simplicity, we omit the COMP algorithm, as EOMP---which has a similar power saving mechanism---almost always achieves lower complexity for a given performance. We propose two types of architectures: (i) fully-unrolled parallel adder-tree (AT)-based architectures which perform one matrix-vector product per clock cycle and (ii) sequential MAC-based architectures which achieve lower throughput but offer greater power savings when compared to the AT-based architectures as detailed in \fref{sec:MACpower}.

\begin{figure}[tp]
	\centering
	\begin{subfigure}[b]{0.38\columnwidth}
		\includegraphics[width=\columnwidth]{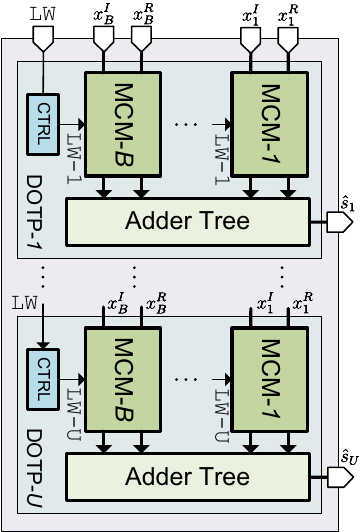}
		\caption{AT-based equalizer}
		\label{fig:MVMa}
	\end{subfigure}
	\hfill
	\begin{subfigure}[b]{0.58\columnwidth}
		\includegraphics[width=\columnwidth]{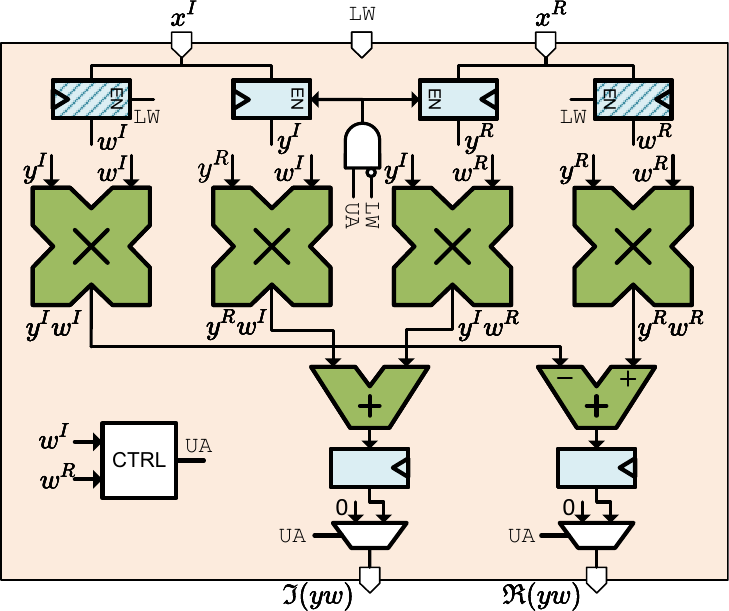}
		\caption{MCM architecture}
		\label{fig:MCM}
	\end{subfigure}
	
	\caption{Architecture of (a) adder-tree (AT)-based equalizer for ALMMSE and EOMP and (b) mute-capable complex multiplier (MCM).}
	\label{fig:MVM}
	
\end{figure}

\subsection{AT-EOMP Architecture} \label{sec:vlsi:LMMSE}

\begin{figure}[tp]
	\centering
	\begin{subfigure}[t]{0.9\columnwidth}
		\includegraphics[width=\columnwidth]{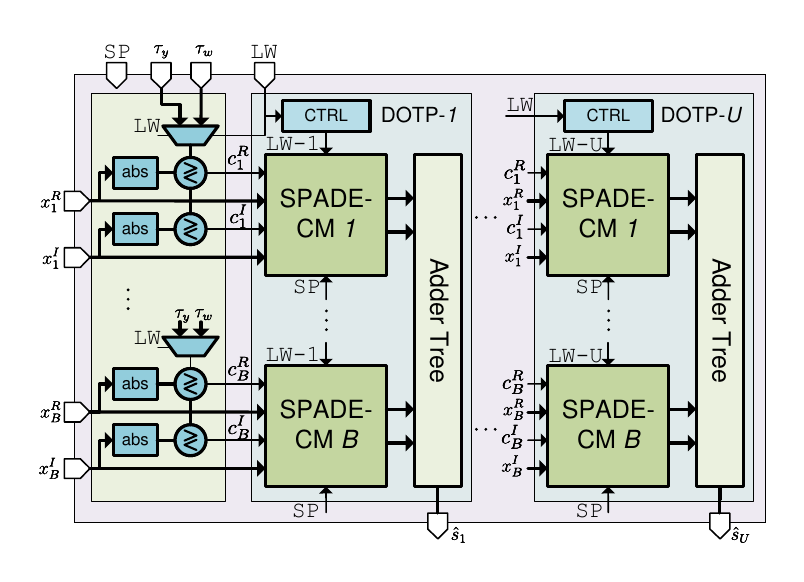}
		\vspace{-0.8cm}
		\caption{AT-SPADE architecture}
		\label{fig:SPMVMa}
	\end{subfigure}
	\vspace{0.3cm}
	
	\begin{subfigure}[t]{0.9\columnwidth}
		\includegraphics[width=\columnwidth]{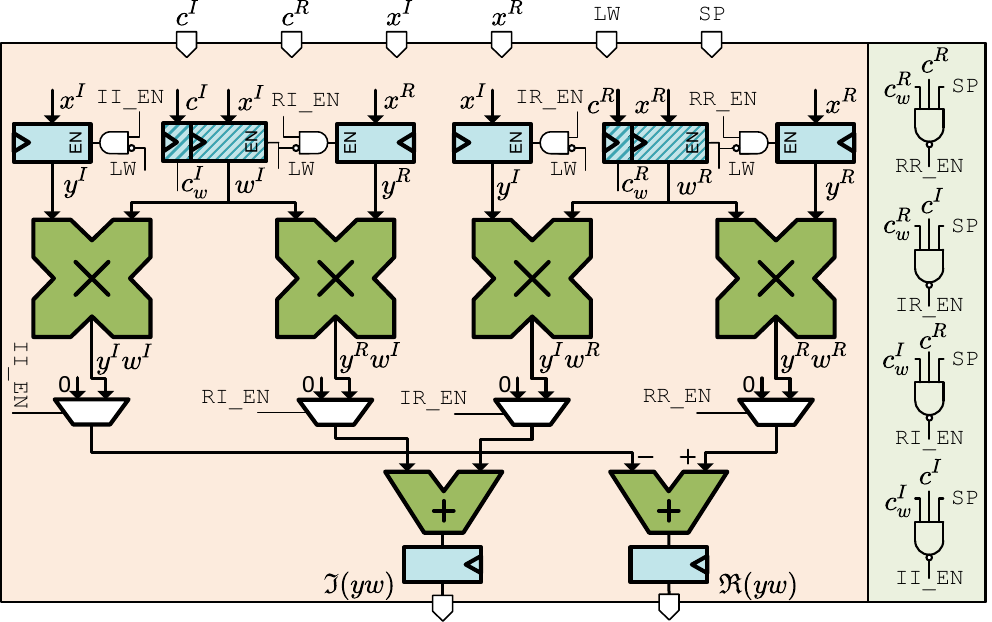}
		\caption{SPADE-CM architecture}
		\label{fig:SPCM}
	\end{subfigure}
	
	%\vspace{-0.2cm}
	\caption{Architecture of AT-SPADE and SPADE-CM equalizers.}
	\label{fig:MVM_SP}
\end{figure}

\begin{figure}[tp]
	\centering
	\begin{subfigure}[t]{0.9\columnwidth}
		\includegraphics[width=\columnwidth]{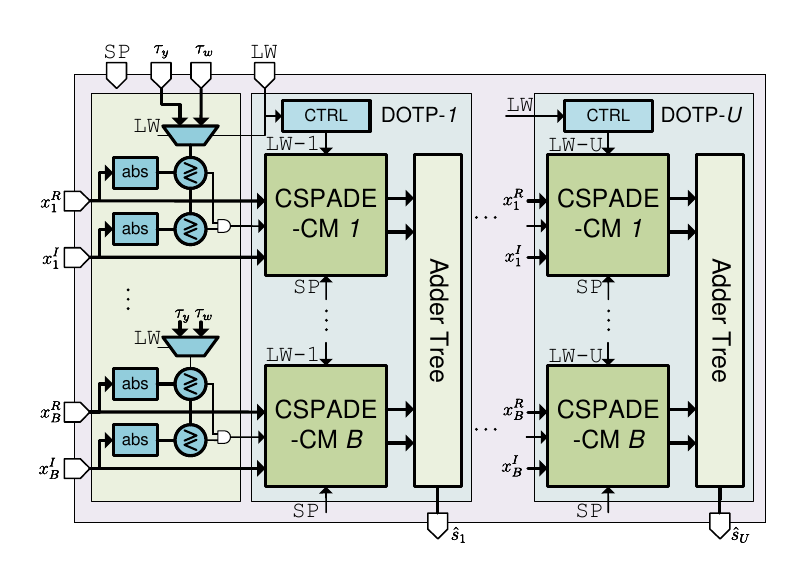}
		\vspace{-0.8cm}
		\caption{AT-CSPADE architecture}
		\label{fig:CSPMVMa}
	\end{subfigure}
	\vspace{0.2cm}
	
	\begin{subfigure}[t]{0.9\columnwidth}
		\includegraphics[width=\columnwidth]{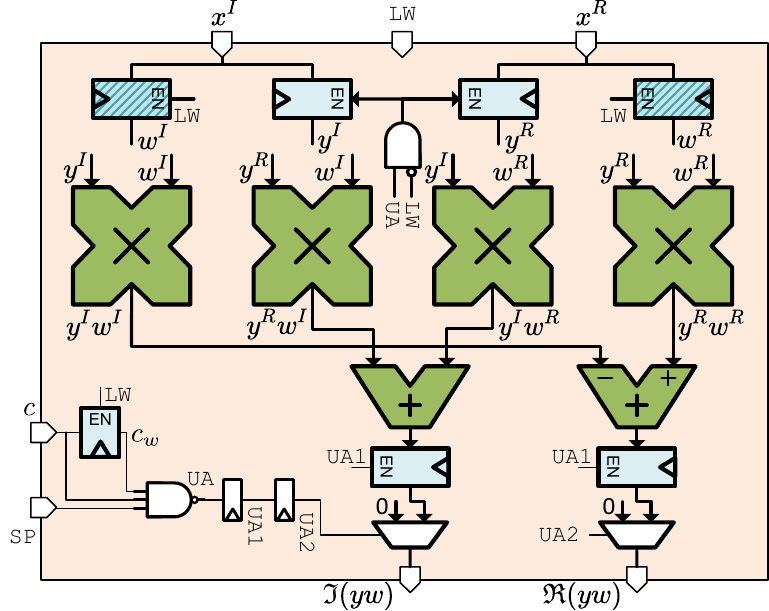}
		\caption{CSPADE-CM architecture}
		\label{fig:CSPCM}
	\end{subfigure}
	
	\caption{Architecture of AT-CSPADE and CSPADE-CM.}
	\label{fig:CSPADE_MVM}
\end{figure}

The parallel AT-based EOMP equalizer is a fully unrolled $U \times B$ matrix-vector multiplier as shown in \fref{fig:MVMa}
consisting of $U$ dot product (DOTP) modules. Each DOTP module consists of~$B$ \emph{mute-capable} complex multipliers (MCM) and an internally pipelined adder tree to sum up the $B$ partial products. 
The internal architecture of the MCM is detailed in \fref{fig:MCM}, which consists of four real-valued multipliers and two adders to perform the complex multiplication, as well as pipeline registers and some control circuitry. The MCM has three main input ports: $x^R$ and $x^I$ for the real and imaginary components of the input signal, respectively, and a load weight (LW) signal. Whenever LW is asserted, the inputs are interpreted as equalization weights and stored as $w^R$ and $w^I$ in the weight registers; whenever LW is deasserted, they get captured in the $y^R$ and $y^I$ registers, provided that at least one of $w^R$ and $w^I$ is nonzero; i.e., when the unit is active, indicated by the UA signal. 
The $2B$ input signals ($x^R_1, x^I_1, \ldots, x^R_B, x^I_B$) of the equalizer are routed to all DOTP modules, and carry either a row of the EOMP equalization matrix ${\bW}$ or the received vector ${\bmy}$, depending on the load weight (LW) signal. To load new equalization matrix into the equalizer, LW is asserted high for $U$ clock cycles, during which the rows of the equalization matrix are input through the $2B$ ports one after the other. 
Within each MCM, if both the real and imaginary parts of the weight are zero, then that MCM becomes inactive and mutes its registers to save dynamic power. In this case, it outputs $0 + j0$ using the multiplexers.
In all considered AT-based architectures, one matrix vector multiplication is performed per clock cycle, resulting in a data detection throughput of $UQf_\text{clk}$ bits per second, where $f_\text{clk}$ is the clock frequency in Hertz and $Q$ is the number of bits per constellation symbol.

\subsection{AT-ALMMSE architecture}

The AT-based ALMMSE equalizer uses the same architecture as in \fref{fig:MVMa}. Rather than the EOMP equalization matrix, the antenna-domain LMMSE equalization matrix $\bar{\bW}$ is provided as input.  The bitwidths of the fixed-point signals in ALMMSE are different from those of EOMP (and the other beamspace equalizers), as discussed in \fref{sec:fixp}. Therefore, the area of the ALMMSE equalizer is smaller than that of the beamspace equalizers, as shown in \fref{sec:area_brk}.

\subsection{AT-SPADE Architecture} \label{sec:SPADE_arch}
The architecture used for AT-based  SPADE  is depicted in \fref{fig:SPMVMa}. The DOTP modules are built from a special type of complex-valued multiplier (CM) referred to as SPADE-CM, shown in \fref{fig:SPCM}. As discussed in \fref{sec:spade}, each real-valued multiplier in SPADE is active only when at least one of its operands has an absolute value larger than the corresponding threshold. To avoid redundant comparisons within SPADE-CMs, AT-SPADE contains circuitry that computes the absolute values of the real and imaginary parts of the input signals and compares them with the threshold $\tau_w$ or $\tau_y$ depending on the LW signal; if LW is high, then the inputs carry the equalization matrix entries, so the threshold is $\tau_w$, otherwise $\tau_y$. The resulting comparison bits are then broadcast to all DOTP modules and fed into the SPADE-CMs through the $c^R$ and $c^I$ ports, shown in \fref{fig:SPCM}. 
\subsubsection*{Muting Mechanism of SPADE-CM}
The control circuitry shown on the right of \fref{fig:SPCM} controls which registers inside SPADE-CM are enabled. 
When the save power (SP) signal is asserted, SPADE-CM saves dynamic power by disabling (freezing) the input registers based on the incoming comparison bits $c^R$ and $c^I$ and the weight comparison bits $c_w^R$ and $c_w^I$ stored during the weight loading phase. 

When SP is deasserted, no muting is performed and SPADE-CM operates like a normal complex multiplier. The mute-capable area of each SPADE-CM consists mainly of the four~$y$ registers (i.e., registers whose output is $y^R$ or $y^I$) plus the four real-valued multipliers, which accounts for approximately $77\%$ of the area of the SPADE-CM. Note that the MUXes after the multipliers, the adders and the output registers are not mute-capable in SPADE-CM. The area breakdown of the $64 \times 8$ matrix-vector multiplier of AT-SPADE in \fref{sec:area_brk}  reveals that $90\%$ of the overall equalizer area is due to SPADE-CMs and the rest are due to the adder trees and other circuitry. Hence, $70\%$ of the total area of the AT-SPADE is mute-capable.

\subsection{AT-CSPADE Architecture} \label{sec:CSPADE_arch}
The AT-CSPADE architecture depicted in \fref{fig:CSPMVMa}, is similar to AT-SPADE. The differences between AT-CSPADE and AT-SPADE are (i) the threshold computation circuitry, which implements the complex comparison detailed in \fref{sec:cspade} and, (ii) DTOPs in CSPADE are built from CSPADE-CMs, whose internal architecture is shown in \fref{fig:CSPCM}. 
\subsubsection*{Key Differences of SPADE-CM and CSPADE-CM} 
There are two subtle differences between the SPADE-CM and CSPADE-CM architectures shown in \fref{fig:SPCM} and \fref{fig:CSPCM}, respectively, which cause the difference between the area and power of SPADE and CSPADE equalizers in \fref{sec:impl_res}.
\begin{itemize}
	\item  SPADE-CM has $6$ input registers; $2$ for storing the real and imaginary parts of the weight ($w^R$ and $w^I$), and $4$ for the real and imaginary parts of the received input ($y^R$ and $y^I$). The reason why SPADE-CM needs two registers for each real/imaginary part of $y$, is to be able to independently mute each of the four real-valued multipliers.
	CSPADE-CM, however, contains only $4$ input registers---similar to the plain complex multiplier of \fref{fig:MCM}---since in CSPADE-CM, either the entire complex multiplier is active or muted.
	\item To support independent muting of the real-valued multipliers, SPADE-CM additionally requires a muting MUX at the output of each real-valued multiplier, and hence, requires $4$ such MUXes, while CSPADE-CM requires only $2$ muting MUXes at the output of the complex multiplier. Furthermore, it is possible to put these two muting MUXes after the output pipeline registers of CSPADE-CM; by doing so we include the output registers in the mute-capable area of CSPADE-CM. Note that this is beneficial for power saving, since the area of the output registers are around $5.5\times$ larger than the two MUXes. Therefore, the mute-capable area of CSPADE-CM consists of the the two $y$ registers ($w$ registers are disabled anyways when $\text{LW}=0$), the four real-valued multipliers, the two adders and the two output registers, which stand for about $92\%$ of the total area of CSPADE-CM, while for SPADE-CM, the mute-capable area is about $77\%$. 
	The area breakdown in \fref{sec:area_brk} for the $64 \times 8$ matrix-vector multiplier of AT-CSPADE shows that $89\%$ of the overall equalizer area is due to CSPADE-CMs. Hence, $83\%$ of the total area of the AT-CSPADE is mute-capable, while for AT-SPADE the mute-capable area is $70\%$.
\end{itemize}

\subsection{MAC-Based Architecture} \label{sec:CSPADE_MAC}

\begin{figure}[tp]
	\centering
	\includegraphics[width=0.8\columnwidth]{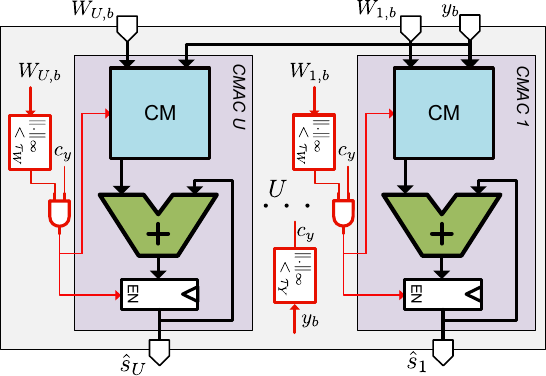}
	\caption{The architecture of MAC-CSPADE consisting of $U$ complex MACs. The circuitry marked with red perform the CSPADE thresholding and enable/disable the CMACs accordingly. The MAC-ALMMSE design uses the same architecture, but without the red circuitry.}
	\label{fig:CSPADE_MAC}
\end{figure}

So far we have only considered fully unrolled AT-based architectures. The AT-based architectures achieve extremely high throughput (i.e., one matrix-vector product per clock cycle) and are suitable for scenarios where many equalization tasks are performed per preprocessing operation. A drawback of this architecture---as will be discussed in \fref{sec:pwr_brk}---is that it does not harvest the full potential of SPADE/CSPADE, since the adder tree is excluded from the mute-capable circuitry. 

We now present an alternative architecture for ALMMSE and CSPADE equalization which is based on complex MAC units, as shown in \fref{fig:CSPADE_MAC}. This architecture performs one matrix-vector product $\hat{\bms} = \bW \bmy$ every $B$ clock cycles using $U$ parallel complex MACs (CMACs), each performing one of the dot products in $\bW \bmy$.  
The circuitry marked with red in \fref{fig:CSPADE_MAC} performs the CSPADE thresholding and activates the CMACs based on CSPADE technique. \fref{fig:CSPADE_MAC} depicts the MAC-CSPADE architecture; the MAC-based ALMMSE equalizer has the same architecture, excluding the CSPADE thresholding circuitry. 
MAC-CSPADE offers more power savings compared to AT-CSPADE. The reason is that whenever a partial product in the dot product is to be skipped using the CSPADE technique in the MAC-based architecture, we can mute the entire CMAC and save power in both the multiplier and the accumulator, as opposed to the AT-based architecture, in which only the multipliers can be muted. Furthermore, the CMs inside the MAC-CSPADE are plain complex-valued multipliers without any additional MUXes at the output to produce $0 + j0$ when skipping a complex multiplication; when the CMAC freezes for a certain partial product, that partial product is simply not accumulated to the result of the dot product.

\section{VLSI Implementation results} \label{sec:impl_res}

In order to (i) evaluate the efficacy of the proposed beamspace equalization schemes in practice and (ii) compare them with antenna-domain equalization, we implemented ALMMSE, EOMP, SPADE, and CSPADE equalizer designs for a  $64 \times 8$ system, based on the VLSI architectures discussed in \fref{sec:vlsi}.
In this section we present the post-layout VLSI implementation results in a 22\,nm FDSOI process, simulated with a supply voltage of $0.8$\,V at room temperature. We emphasize that the BER performance of the implemented fixed-point designs correspond to those shown in~\fref{fig:BER:a} and \fref{fig:BER:b} for LoS and non-LoS channels, respectively.

\subsection{Area Breakdown} \label{sec:area_brk}

\begin{figure}[tp]
	\centering
	\includegraphics[width=0.75\linewidth]{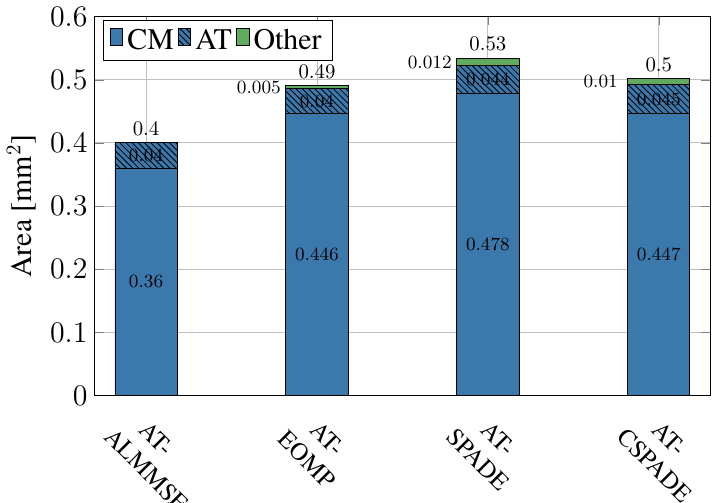}
	\vspace{-0.1cm}
	\caption{Post-layout area breakdown of AT-based equalizers for a $64\times 8$ system}
	\label{fig:area_brk}
\end{figure}

\fref{fig:area_brk} shows the area breakdown of the AT-based equalizers into their main modules. The plain blue part of each bar shows the area of the complex multipliers (CM), while the hatched blue part shows the area of the adder trees (ATs). The green part shows the area of miscellaneous logic, which contain different circuitry for each design, including the threshold comparison circuitry for AT-SPADE and AT-CSPADE.   

\fref{fig:area_brk} shows that the AT-ALMMSE equalizer has a smaller area than their beamspace counterparts. The reason is that the antenna-domain inputs have smaller bitwidths compared to beamspace inputs, as discussed in \fref{sec:fixp}. We also notice that most of the equalizer area is due to the complex-valued multipliers, which occupy around $90$\% of the total area.
It is interesting to note that the area of AT-CSPADE is $6\%$ smaller than AT-SPADE, which is mainly due to the fact that CSPADE-CMs are smaller than SPADE-CMs; due to the architectural differences detailed in \fref{sec:CSPADE_arch}.
The area of the ALMMSE equalizer is smaller than CSPADE equalizer in the  MAC-based designs as well: the area of MAC-ALMMSE and MAC-CSPADE are $6\,000\,\upmu \textnormal{m}^2$ and $7\,887\,\upmu \textnormal{m}^2$, respectively.

\subsection{Power Breakdown} \label{sec:pwr_brk}

\begin{figure}[tp]
	\centering
	\begin{subfigure}[t]{0.75\linewidth}
		\includegraphics[width=\linewidth]{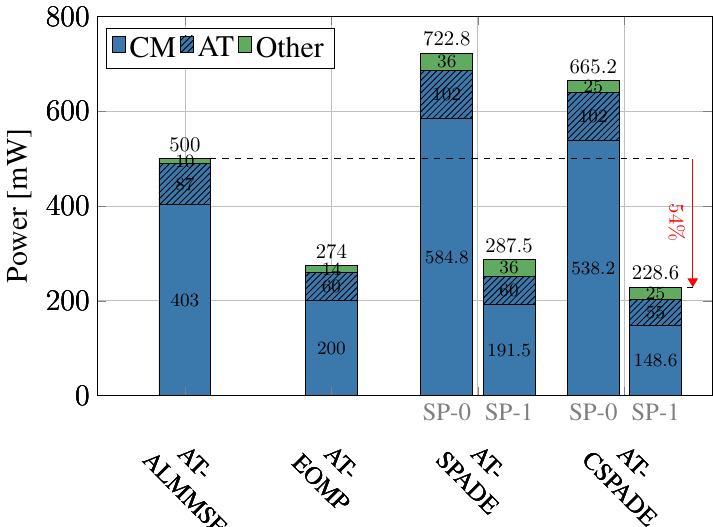}
		\hspace{-1cm}
		\caption{LoS} 
		\label{fig:pwr_brk_64x8_LOS}
	\end{subfigure}
	\hfill
	\begin{subfigure}[t]{0.75\linewidth}
		\includegraphics[width=\linewidth]{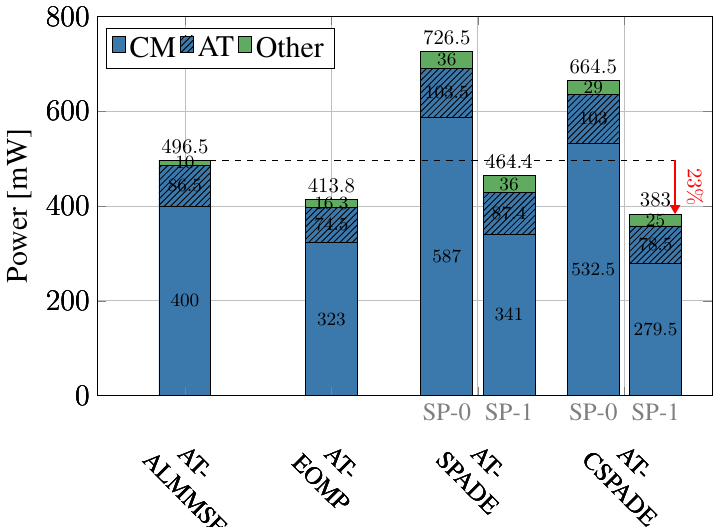}
		\caption{non-LoS}
		\label{fig:pwr_brk_64x8_NLOS}
	\end{subfigure}
	
	\vspace{-0.1cm}
	\caption{Power breakdown of AT-based equalizers with LoS (a) and non-LoS (b) channels. For AT-SPADE and AT-CSPADE, we show both the power breakdown both with power saving (SP-1) and without power saving (SP-0).}
	\label{fig:pwr_brk}
\end{figure}

\fref{fig:pwr_brk} shows a breakdown of the power consumption of the AT-based equalizers using post-layout power simulations, with the node switching activity extracted from stimuli generated with LoS and non-LoS channels at an SNR of $10$\,dB. 
In the power simulations for EOMP, the $\bW$ inputs are generated with the density coefficients $\delta$ given in \fref{fig:BER:a} and \fref{fig:BER:b}, for LoS and non-LoS, respectively. Similarly, for SPADE and CSPADE, the threshold pairs $(\tau_y, \tau_w)$ correspond to the multiplier activity rates $\alpha$ in \fref{fig:BER:a} and \fref{fig:BER:b}.

This power breakdown shows the power consumption due to complex multipliers, adder trees, and miscellaneous circuitry. For SPADE and CSPADE, we show the power when save power (SP) signal is asserted (labeled as SP-1) and when it is deasserted (labeled as SP-0). We observe that when SP is asserted, AT-CSPADE offers $54\%$ and $23\%$ power savings in LoS and non-LoS channels compared to \mbox{AT-ALMMSE}, respectively. AT-SPADE offers 
$42\%$ and $6\%$ power savings in LoS and non-LoS channels compared to \mbox{AT-ALMMSE} equalizer, respectively.

\subsubsection{Power Savings vs. Multiplier Activity Rate} \label{sec:CSPADEMVM_power}

For SPADE and CSPADE, the multiplier activity rate $\alpha$ represents the average ratio of active multiplications to the total number of multiplications involved in the equalization (cf. \fref{sec:activity_rate}). Therefore, $\alpha$ is expected to estimate the power savings achieved by activating SPADE/CSPADE. 
Note, however, that $\alpha$ only represents the ratio of active multiplications and ignores additions. Therefore,  $\alpha$ can only predict the power savings achieved in the multiplier circuitry. 
In the CSPADE results presented in \fref{fig:pwr_brk}, $\alpha = 0.21$ for LoS and $\alpha = 0.45$ for non-LoS stimuli, as given in \fref{fig:BER:a} and \fref{fig:BER:b}, respectively. In the following, we analyze the power of CSPADE only; the same arguments apply to SPADE as well.

As we see in \fref{fig:pwr_brk}, the ratio of power of CSPADE-CMs with power savings activated (SP-1) to the power of CSPADE-CMs without power savings activated (SP-0) is indeed close to $\alpha$, but slightly higher, which is due to the fact that not all of the area of CSPADE-CMs is mute-capable. As discussed in \fref{sec:CSPADE_arch}, $92\%$ of CSPADE-CM area is mute-capable. Furthermore, when a CSPADE-CM is active, not all of its internal nodes are toggling; the switching rate---which determines the dynamic power---depends on the instantaneous inputs. In the estimate using $\alpha$, however, an active multiplier is assumed to have a switching rate of $1$ and a muted multiplier to have a switching rate of $0$, which is an oversimplification of the real situation. 

To analyze the power of the entire AT-CSPADE equalizer (including both complex multipliers and the adder trees), we note that as discussed in \fref{sec:CSPADE_arch}, only $83\%$ of the area of the AT-CSPADE equalizer is mute-capable. Hence, with multiplier activity rate of $\alpha = 0.21$ in LoS and $\alpha = 0.45$, in non-LoS, CSPADE can potentially reduce the LoS power by at most $0.17 + 0.21 \times 0.83 = 0.34$ times, i.e., from $665$\,mW down to $228.7$\,mW, and the non-LoS power by at most $0.17 + 0.45 \times 0.83 = 0.54$ times, i.e., from $665$\,mW down to $361$\,mW. These power estimates are very close to the actual simulated power consumption shown in \fref{fig:pwr_brk}.

\subsubsection{SPADE vs. CSPADE}
A key observation in \fref{fig:pwr_brk} is that the power of AT-CSPADE with SP-1 is around $15\%$ lower than the power of AT-SPADE with SP-1, even though AT-SPADE is operated with a smaller multiplier activity rate than AT-CSPADE in both LoS and non-LoS channels, i.e., $\alpha_\text{SPADE} = 0.15$ in LoS and $\alpha_\text{SPADE} = 0.43$ in non-LoS, while $\alpha_\text{CSPADE} = 0.21$ in LoS and $\alpha_\text{CSPADE} = 0.45$ in non-LoS. The lower power of AT-CSPADE is due to the following reasons: 
\begin{itemize}
	\item As discussed in \fref{sec:CSPADE_arch}, the area of AT-CSPADE is around $6\%$ smaller than AT-SPADE, since CSPADE-CMs are smaller than SPADE-CMs.
	\item As detailed in \fref{sec:SPADE_arch} and \fref{sec:CSPADE_arch}, for AT-SPADE equalizer, $70\%$ of the total area is mute-capable, while for AT-CSPADE equalizer, $83\%$ is mute-capable.
\end{itemize}

\subsubsection{MAC-Based Architecture} \label{sec:MACpower}
The above discussions reveal that the AT-based architecture does not harvest the full potential of SPADE/CSPADE for power saving, since the adder tree is not part of the mute-capable circuitry.
The MAC-based architecture for CSPADE presented in \fref{sec:CSPADE_MAC} is expected to achieve higher power savings, using the same threshold pair and stimuli used in the AT-CSPADE.

\begin{figure}[tp]
	\centering
	\begin{subfigure}[t]{0.48\linewidth}
		\includegraphics[width=\linewidth]{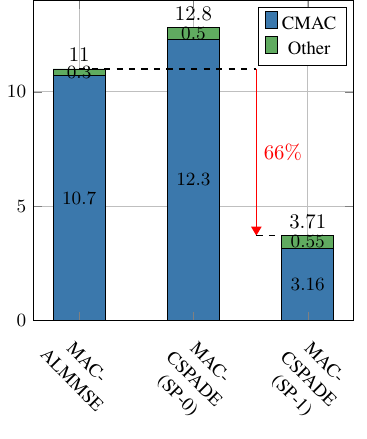}
		\hspace{-1cm}
		\caption{LoS} 
		\label{fig:pwr_brk_MAC:LOS}
	\end{subfigure}
	\hfill
	\begin{subfigure}[t]{0.48\linewidth}
		\includegraphics[width=\linewidth]{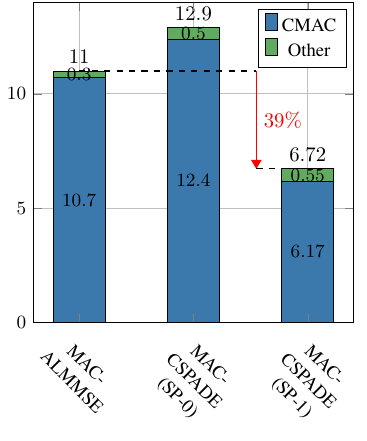}
		\caption{non-LoS}
		\label{fig:pwr_brk_MAC:NLOS}
	\end{subfigure}
	
	\vspace{-0.1cm}
	\caption{Power consumption of MAC-ALMMSE and MAC-CSPADE with $U=8$ CMACs in LoS (a) and non-LoS (b) channels. For MAC-CSPADE we show both the power breakdown both with power saving (SP-1) and without power saving (SP-0).}
	\label{fig:pwr_brk_MAC}
\end{figure}

\fref{fig:pwr_brk_MAC} shows the power consumption of MAC-ALMMSE and MAC-CSPADE made of $U=8$ complex MACs, in LoS and non-LoS channels. We observe that MAC-CSPADE with \mbox{SP-1} achieves $71\%$ and $48\%$ power savings compared to \mbox{SP-0} in LoS and non-LoS channels, which is larger than the power savings achieved in AT-CSPADE with SP-1 compared to SP-0 (cf. \fref{sec:CSPADEMVM_power}). Furthermore, MAC-CSPADE with \mbox{SP-1} achieves $66\%$ and $39\%$ power savings compared MAC-ALMMSE, which is higher than the corresponding power savings in the AT-based architectures.

\begin{table*}[tp]
	\centering
	\renewcommand{\arraystretch}{1.1}
	\renewcommand{\thefootnote}{\alph{footnote}}
	\begin{minipage}[c]{1\textwidth}
		\centering
		\caption{Comparison with the state-of-the-art massive MU-MIMO data detectors.}
		\label{tbl:copmarison}
		\resizebox{0.99\textwidth}{!}{%			
			\begin{tabular}{@{}lcc|ccccc@{}}
				\toprule
				~ & \multicolumn{2}{c|}{This work} & \cite{sirpac21} & \cite{liu20} & \cite{tang21} & \cite{prabhu17} & \cite{Mahdavi20} \\
				\midrule
				Algorithm  & ALMMSE & CSPADE & LMMSE & RCG  & MPD  & LMMSE & LMMSE-SB\\
				System dimension $B\times U$ & \multicolumn{2}{c|}{$64\times 8$} & $32\times 16$ & $128\times 8$ & $128\times 32$  &  $128\times 8$\ & $128\times 16$\\
				Modulation [QAM] & \multicolumn{2}{c|}{16} & 16 & 64 & 256 & 256 & 16\\
				
				Includes preprocessing & \multicolumn{2}{c|}{no} & no & yes \footnotemark[1] & no  & yes  \footnotemark[1] & yes  \footnotemark[1]\\
				%SNR loss [dB]\footnotemark[1] & 0.7 & 0.5 & 0.9 & error flow & NA & better than LMMSE & \\
				\midrule
				Technology [nm] &  \multicolumn{2}{c|}{22} & 65 & 65 & 40  & 28 & 28\\
				Core voltage [V] &  \multicolumn{2}{c|}{0.8} & 1.0 & 1.2 & 0.9  & 0.9 & NA \\
				Core area [$\text{mm}^2$] & 0.4 & 0.5  &  2.41 & 1.6 & 0.58 & -- \footnotemark[2] \footnotemark[3] & -- \footnotemark[3]\\
				Maximum clock frequency [MHz] & 1\,000 & 1\,000 & 312 & 500 & 425 & 300 & 560\\
				
				Maximum throughput [Gbps] \footnotemark[4] &32  & 32 & 9.98 & 1.0 & 1.38 & 0.15 & 2.24\\
				Power [mW] & 500 & 229 \footnotemark[5]   & 290 & 120 & 220.6  & 18 & 251 \\
				\midrule
				Energy efficiency [pJ/b] & 15.6 & 7.2 & 29 & 120 & 160 & 120 & 112\\
				Area efficiency [Gbps/$\text{mm}^2$] & 80 & 64 & 4.15 & 0.63 & 2.38 & 1.25 & --\\
				\midrule
				Norm. energy efficiency [pJ/b]\footnotemark[6]\footnotemark[7] & 15.6 & 7.2  & 12.5 & 18 &  17.3  & 74.5 & 34.8 \footnotemark[8]\\
				Norm. area efficiency [Gbps/$\text{mm}^2$] \footnotemark[6]\footnotemark[7] & 80 & 64 & 53 & 16.1 & 57.2 & -- & --\\
				\bottomrule
			\end{tabular}
			\newcommand{\footmark}[1]{$^{\footnotesize \textnormal{#1}}$}
			\footnotetext{\noindent%
				\footmark{a}although these designs include preprocessing circuitry, the reported throughput and power in \cite{prabhu17} and \cite{Mahdavi20} exclude preprocessing, and for \cite{liu20} we took the implementation results without preprocessing; 
				\footmark{b}not reported information indicated by -- ; 
				\footmark{c}for the designs \cite{prabhu17} and \cite{Mahdavi20}, only equivalent gate count is reported;
				\footmark{d}computed with the maximum  reported clock frequency for $16$-QAM in all designs to isolate the effect of modulation order;
				\footmark{e}computed with node switching activity extracted from stimuli generated with LoS channels;
				\footmark{f}technology normalized to $22$\,nm at $0.8$\,V nominal core voltage, given the assumptions: $f_\text{clk}\sim s$, $\mathrm{area}\sim 1/s^2$ and $\mathrm{power}\sim 1/(V^2)$, where $s$ is the ratio of technology nodes and $V$ is the ratio of core voltages \cite{Rabaey_book};
				\footmark{g}throughput, area and power results are scaled based on the architecture of each reference design individually, according to the explanations in \fref{sec:normalization};
				\footmark{h}we used the same core voltage as \cite{prabhu17} since it has the same technology as \cite{Mahdavi20};
			}
		}		
	\end{minipage}
	
\end{table*}

\subsection{Comparison with the State-of-the-Art} \label{sec:impl_comp}

The results in \fref{sec:area_brk} and \fref{sec:pwr_brk} together with the results from \fref{sec:tradeoff} reveal that CSPADE achieves the best performance-power trade-off among the proposed beamspace equalizers (both with AT-based and MAC-based architectures), as it consumes the lowest power among other equalizers, with virtually the same error-rate performance. Furthermore, this approach requires the same preprocessing complexity as ALMMSE, in contrast to EOMP and COMP which involve complicated preprocessing. Therefore, we now focus on the post-layout implementation results of AT-CSPADE as the representative of the proposed beamspace equalizers.

In order to evaluate the efficiency of the proposed designs, we summarize the key post-layout implementation results of AT-ALMMSE and AT-CSPADE equalizers for a $64 \times 8$ system, as well as several state-of-the-art massive MU-MIMO detectors in \fref{tbl:copmarison}. For CSPADE, we report the power with LoS stimuli. The energy efficiency is defined as the ratio of power divided by the throughput (in pJ/b), and the area efficiency as the ratio of throughput achieved per millimeters squared of area.
To enable a fair comparison of the energy and area efficiencies reported in \fref{tbl:copmarison}, we project the results of the reference designs onto a $22$\,nm process using the approximations given in the footnote of \fref{tbl:copmarison}. Additionally, since the reference designs are implemented for different system dimensions than our design, we normalize their area, power, and throughput to a system with $B = 64$ antennas and $U = 8$ UEs as follows:

\subsubsection{System Dimension Normalization} \label{sec:normalization}

For the design \cite{prabhu17}, $B=128$ only affects computation of the Gram matrix and matched filter (MF), which are excluded from the reported results. The reported throughput for \cite{prabhu17}, only includes the forward and backward substitution operation given that MF, Gram matrix and its Cholesky decomposition are precomputed.  
The forward and backward substitution modules are implemented using systolic arrays with $U$ processing elements, with a throughput of one symbol per $8$ clock cycles, as detailed in~\cite{prabhu_phd}.

For the design in \cite{Mahdavi20}, the reported throughput is $4$ times the clock frequency, which corresponds to one $16$-QAM symbol per clock cycle.
This indicates that the reported throughput does not include preprocessing nor MF computation and only takes into account forward and backward substitution operations which require at least $U$ cycles to produce $U$ symbol estimates. We emphasize that the architecture presented in \cite{Mahdavi20} and \cite{Mahdavi19} would require at least $M'$ cycles to perform MF, where $M'=80$ is the number of selected beams in the LMMSE-SB algorithm.
Since the energy efficiency in \cite{Mahdavi20} is computed by dividing the reported throughput by the power consumption, we conclude that the reported power also excludes preprocessing and corresponds only to the power of forward and backward substitution, which consists of $3U-2$ general processing elements (PEs). Since $U=16$ in \cite{Mahdavi20}, we scale the reported power by $8/16$. However, we do not scale the throughput number, since changing $U$ does not change the throughput, which remains to be one symbol per cycle by design.

The message passing detector (MPD) in \cite{tang21} does not include preprocessing and assumes that Gram computation and MF are performed externally and provided as input to the design. Therefore, $B$ does not affect the reported results. Their proposed architecture requires $U/2$ Interference-cancellation PEs (IPEs), each consisting of $U$ MACs. Therefore, the area and power of \cite{tang21} scales with $U^2$ and the throughput by $U$. Note that, due to the proposed symbol-hardening scheme, the constellation matching PEs (CPEs) reduce to wiring only and therefore the area and power is independent of the supported constellation. Therefore, we scale the area and power reported in \cite{tang21} by $(8/32)^2$ and its throughput by $(8/32)$.
We note that as reported in \cite{tang21}, the performance of this iterative detector relies on uncorrelateness assumption and channel hardening and even for Rayleigh fading channels exhibits a noticeable performance loss compared to LMMSE equalization---this detector is not suitable for channels that exhibit correlation among UEs.

The recursive conjugate gradient (RCG) detector proposed in \cite{liu20} reports the implementation results with and without preprocessing (i.e., MF and Gram computation); we took the results without preprocessing, for a fair comparison. Therefore, $B$ does not affect the results and $U=8$ as in our design. 

The design in \cite{sirpac21}, is a bit-serial matrix-vector multiplier, whose area and power scales with $BU$ and the throughput with~$U$. Hence, we only scale the reported throughput by $8/U$.

\subsubsection{Comparison Results}
We observe in \fref{tbl:copmarison} that our proposed designs achieve the highest throughput among other detectors thanks to the fully unrolled architecture and internal pipelining which reduces the critical path and allows for high clock frequencies. Furthermore, our designs achieve the best-in-class area- and energy-efficiency.
We note that MPD \cite{tang21} achieves competitive area- and energy-efficiency. However, as noted in \cite{tang21}, MPD relies on the assumption of channel hardening in massive MIMO systems and suffers significant performance degradation compared to LMMSE under correlated channels which are typical in mmWave frequencies~\cite{rappaport15a}. The RCG detector \cite{liu20} also exhibits similar energy-efficiency to our designs. However, the results in \cite{liu20} show that the performance of RCG is worse than LMMSE with Rayleigh fading channels. As shown in \cite{mirfarshbafan24} and \cite{ZixiaoLi25}, RCG incurs even larger performance degradation compared to LMMSE, with realistic LoS and non-LoS mmWave channels, which have some degree of correlation between the UE channels.
Furthermore, as shown in \fref{sec:BER} the design in~\cite{Mahdavi20} (LMMSE-SB), incurs a large performance loss compared to ALMMSE and the beamspace equalizers, particularly in LoS channels.
In contrast, as shown in \fref{fig:SNRop2:a} and \fref{fig:SNRop2:b}, CSPADE incurs less than $0.6$ dB SNR loss at $\text{BER} = 10^{-3}$ compared to LMMSE, while the gap is smaller at higher target BERs (e.g., $\text{BER} = 10^{-2}$), as observed in \fref{fig:BER}.

\section{Conclusions} \label{sec:conclusion}
We have discussed existing beamspace data detection algorithms and proposed CSPADE, a novel improved variant of SPADE. 
We have evaluated the performance-complexity trade-offs of these algorithms using fixed-point simulations. Furthermore, we have identified that beamspace signals exhibit larger dynamic range which calls for larger bitwidth in their fixed-point representation.
We have proposed VLSI architectures for SPADE, CSPADE, EOMP, and ALMMSE, and presented corresponding post-layout VLSI implementation power and area results. Based on these results, we have demonstrated that despite the larger bitwidths required for beamspace signals, the proposed CSPADE equalizer offers significant power savings---up to $66\%$---compared to conventional antenna-domain linear equalization. Furthermore, our comparison with the state-of-the-art  massive MU-MIMO equalizers reveals that CSPADE achieves the best-in-class area- and energy-efficiency. Thus, beamspace processing is an effective approach to reduce power consumption of baseband processing in all-digital BS architectures for mmWave massive multiuser MIMO systems. 

\newpage



\begin{thebibliography}{10}
	\providecommand{\url}[1]{#1}
	\csname url@samestyle\endcsname
	\providecommand{\newblock}{\relax}
	\providecommand{\bibinfo}[2]{#2}
	\providecommand{\BIBentrySTDinterwordspacing}{\spaceskip=0pt\relax}
	\providecommand{\BIBentryALTinterwordstretchfactor}{4}
	\providecommand{\BIBentryALTinterwordspacing}{\spaceskip=\fontdimen2\font plus
		\BIBentryALTinterwordstretchfactor\fontdimen3\font minus
		\fontdimen4\font\relax}
	\providecommand{\BIBforeignlanguage}[2]{{%
			\expandafter\ifx\csname l@#1\endcsname\relax
			\typeout{** WARNING: IEEEtran.bst: No hyphenation pattern has been}%
			\typeout{** loaded for the language `#1'. Using the pattern for}%
			\typeout{** the default language instead.}%
			\else
			\language=\csname l@#1\endcsname
			\fi
			#2}}
	\providecommand{\BIBdecl}{\relax}
	\BIBdecl
	
	\bibitem{mirfarshbafan21b}
	S.~H. Mirfarshbafan and C.~Studer, ``{SPADE: Sparsity-Adaptive Equalization}
	for {MMwave} massive {MU-MIMO},'' in \emph{Proc. IEEE Workshop Stat. Signal
		Process. (SSP)}, Aug. 2021, pp. 211--215.
	
	\bibitem{SeyedHadi20b}
	S.~H. {Mirfarshbafan} and C.~{Studer}, ``Sparse beamspace equalization for
	massive {MU-MIMO} {mmWave} systems,'' in \emph{Proc. IEEE Int. Conf. Acoust.,
		Speech, Signal Process. (ICASSP)}, May 2020, pp. 1773--1777.
	
	\bibitem{3gpp22}
	3GPP, ``{5G}; {NR}; base station ({BS}) radio transmission and reception,''
	Apr. 2022, {TS} 38.104 version 16.11.0 Rel.~16.
	
	\bibitem{rappaport15a}
	T.~S. Rappaport, R.~W. {Heath Jr.}, R.~C. Daniels, and J.~N. Murdock,
	\emph{Millimeter Wave Wireless Communications}.\hskip 1em plus 0.5em minus
	0.4em\relax Prentice Hall, 2015.
	
	\bibitem{larsson14a}
	E.~G. Larsson, F.~Tufvesson, O.~Edfors, and T.~L. Marzetta, ``Massive {MIMO}
	for next generation wireless systems,'' \emph{{IEEE} Commun. Mag.}, vol.~52,
	no.~2, pp. 186--195, Feb. 2014.
	
	\bibitem{sohrabi16}
	F.~{Sohrabi} and W.~{Yu}, ``Hybrid digital and analog beamforming design for
	large-scale antenna arrays,'' \emph{{IEEE} J. Sel. Topics Signal Process.},
	vol.~10, no.~3, pp. 501--513, Apr. 2016.
	
	\bibitem{heath-jr.15a}
	R.~W. {Heath Jr.}, N.~Gonz{\'a}lez~Prelcic, S.~Rangan, W.~Roh, and A.~Sayeed,
	``An overview of signal processing techniques for millimeter wave {MIMO}
	systems,'' \emph{{IEEE} J. Sel. Topics Signal Process.}, vol.~10, no.~3, pp.
	436--453, Feb. 2016.
	
	\bibitem{yan2019performance}
	H.~Yan, S.~Ramesh, T.~Gallagher, C.~Ling, and D.~Cabric, ``Performance, power,
	and area design trade-offs in millimeter-wave transmitter beamforming
	architectures,'' \emph{{IEEE} Circuits Syst. Mag.}, vol.~19, no.~2, pp.
	33--58, May 2019.
	
	\bibitem{panagiotis20}
	P.~{Skrimponis}, S.~{Dutta}, M.~{Mezzavilla}, S.~{Rangan}, S.~H.
	{Mirfarshbafan}, C.~{Studer}, J.~{Buckwalter}, and M.~{Rodwell}, ``Power
	consumption analysis for mobile mmwave and sub-{TH}z receivers,'' in
	\emph{2020 IEEE 6G Wireless Summit}, Mar. 2020, pp. 1--5.
	
	\bibitem{schniter14a}
	P.~Schniter and A.~Sayeed, ``Channel estimation and precoder design for
	millimeter-wave communications: {The} sparse way,'' in \emph{Proc. Asilomar
		Conf. Signals, Syst., Comput.}, Nov. 2014, pp. 273--277.
	
	\bibitem{mirfarshbafan19a}
	S.~H. {Mirfarshbafan}, A.~{Gallyas-Sanhueza}, R.~{Ghods}, and C.~{Studer},
	``Beamspace channel estimation for massive {MIMO} {mmWave} systems: Algorithm
	and {VLSI} design,'' \emph{{IEEE} Trans. Circuits Syst. {I}}, pp. 1--14, Sep.
	2020.
	
	\bibitem{Abdelghany19}
	M.~{Abdelghany}, U.~{Madhow}, and A.~{T{\"o}lli}, ``Beamspace local {LMMSE}: An
	efficient digital backend for {mmWave} massive {MIMO},'' in \emph{Proc. IEEE
		Int. Workshop Signal Process. Advances Wireless Commun. (SPAWC)}, Aug. 2019,
	pp. 1--5.
	
	\bibitem{Mahdavi20}
	M.~{Mahdavi}, O.~{Edfors}, V.~{{\"O}wall}, and L.~{Liu}, ``Angular-domain
	massive {MIMO} detection: {Algorithm}, implementation, and design
	tradeoffs,'' \emph{{IEEE} Trans. Circuits Syst. {I}}, vol.~67, no.~6, pp.
	1948--1961, Jan. 2020.
	
	\bibitem{goluntas21}
	E.~Gönülta\c{s}, S.~Taner, A.~Gallyas-Sanhueza, S.~H. Mirfarshbafan, and
	C.~Studer, ``Hardware-aware beamspace precoding for all-digital {mmWave}
	massive {MU-MIMO},'' \emph{{IEEE} Commun. Lett.}, vol.~25, no.~11, pp.
	3709--3713, Aug. 2021.
	
	\bibitem{CAPMIMO}
	J.~{Brady}, N.~{Behdad}, and A.~{Sayeed}, ``Beamspace {MIMO} for
	millimeter-wave communications: System architecture, modeling, analysis, and
	measurements,'' \emph{{IEEE} Trans. Antennas Propag.}, vol.~61, no.~7, pp.
	3814--3827, Jul. 2013.
	
	\bibitem{Song13Beamspace}
	G.~H. {Song}, J.~{Brady}, and A.~{Sayeed}, ``Beamspace {MIMO} transceivers for
	low-complexity and near-optimal communication at mm-wave frequencies,'' in
	\emph{Proc. IEEE Int. Conf. Acoust., Speech, Signal Process. (ICASSP)}, May
	2013, pp. 4394--4398.
	
	\bibitem{SayeedGLOBECOM}
	A.~{Sayeed} and J.~{Brady}, ``Beamspace mimo for high-dimensional multiuser
	communication at millimeter-wave frequencies,'' in \emph{Proc. IEEE Global
		Commun. Conf. (GLOBECOM)}, Dec. 2013, pp. 3679--3684.
	
	\bibitem{GaoNearOptimalBeamspace}
	X.~{Gao}, L.~{Dai}, Z.~{Chen}, Z.~{Wang}, and Z.~{Zhang}, ``Near-optimal beam
	selection for beamspace mmwave massive mimo systems,'' \emph{{IEEE} Commun.
		Lett.}, vol.~20, no.~5, pp. 1054--1057, May 2016.
	
	\bibitem{MahdaviGlobalSIP}
	M.~Mahdavi, O.~Edfors, V.~\"Owall, and L.~Liu, ``A low complexity massive
	{MIMO} detection scheme using angular-domain processing,'' in \emph{IEEE
		Global Conf. on Signal and Inf. Process. (GlobalSIP)}, Feb. 2019, pp.
	181--185.
	
	\bibitem{OMP}
	Y.~Pati, R.~Rezaiifar, and P.~Krishnaprasad, ``Orthogonal matching pursuit:
	recursive function approximation with applications to wavelet
	decomposition,'' in \emph{Proc. Asilomar Conf. Signals, Syst., Comput.}, Nov.
	1993, pp. 40--44 vol.1.
	
	\bibitem{Yoshida22}
	T.~Yoshida, D.~Shirase, T.~Takahashi, S.~Ibi, and S.~Sampei, ``Low-complexity
	large {MIMO} detection based on beam-domain local {LMMSE} filters,'' in
	\emph{Proc. IEEE Int. Conf. Commun. (ICC)}, 2022, pp. 1312--1317.
	
	\bibitem{seethaler10a}
	D.~Seethaler and H.~B{\"o}lcskei, ``Performance and complexity analysis of
	infinity-norm sphere-decoding,'' \emph{{IEEE} Trans. Inf. Theory}, vol.~56,
	no.~3, pp. 1085--1105, Mar. 2010.
	
	\bibitem{risi2014massive}
	C.~Risi, D.~Persson, and E.~G. Larsson, ``Massive {MIMO} with 1-bit {ADC},''
	\emph{arXiv preprint: 1404.7736}, Apr. 2014.
	
	\bibitem{studer16a}
	C.~Studer and G.~Durisi, ``Quantized massive {MU-MIMO-OFDM} uplink,''
	\emph{{IEEE} Trans. Commun.}, vol.~64, no.~6, pp. 2387--2399, Jun. 2016.
	
	\bibitem{mirfarshbafan20}
	S.~H. {Mirfarshbafan}, S.~A. {Nezamalhosseini}, M.~{Shabany}, and C.~{Studer},
	``Algorithm and {VLSI} design for 1-bit data detection in massive
	{MIMO-OFDM},'' \emph{{IEEE} Open J. Circuits Syst. {I}}, vol.~1, pp.
	170--184, Sep. 2020.
	
	\bibitem{sven_thesis}
	S.~Jacobsson, ``Massive multi-antenna communications with low-resolution data
	converters,'' Ph.D. dissertation, Chalmers University of Technology,
	Gothenburg, Sweden, 2019.
	
	\bibitem{tse05a}
	D.~Tse and P.~Viswanath, \emph{Fundamentals of Wireless Communication}.\hskip
	1em plus 0.5em minus 0.4em\relax Cambridge Univ. Press, 2005.
	
	\bibitem{Fang17}
	L.~Fang, H.~Li, D.~David~Huang, and X.~Huang, ``A low cost interpolation based
	detection algorithm for medium-size massive {MIMO-OFDM} systems,'' in
	\emph{International Symp. Comm. Info. Tech. (ISCIT)}, Sep. 2017.
	
	\bibitem{prabhu17}
	H.~Prabhu, J.~N. Rodrigues, L.~Liu, and O.~Edfors, ``A {60pJ/b 300Mb/s} 128×8
	massive {MIMO} precoder-detector in 28nm {FD-SOI},'' in \emph{IEEE Int.
		Solid-State Circuits Conf. (ISSCC) Dig. Tech. Papers}, Mar. 2017, pp. 60--61.
	
	\bibitem{mirfarshbafan24}
	S.~H. Mirfarshbafan and C.~Studer, ``A 46 {Gbps} 12 {pJ/b} sparsity-adaptive
	beamspace equalizer for {mmWave} massive {MIMO} in {22FDX},'' \emph{{IEEE}
		Trans. Circuits Syst. {II}}, Dec. 2024.
	
	\bibitem{Abdelghany20}
	M.~Abdelghany, M.~E. Rasekh, and U.~Madhow, ``Scalable nonlinear multiuser
	detection for {mmWave} massive {MIMO},'' in \emph{Proc. IEEE Int. Workshop
		Signal Process. Advances Wireless Commun. (SPAWC)}, Aug. 2020, pp. 1--5.
	
	\bibitem{knuth1981art2}
	D.~E. Knuth, \emph{The Art of Computer Programming, Volume 2: Seminumerical
		Algorithms}, 2nd~ed.\hskip 1em plus 0.5em minus 0.4em\relax Addison-Wesley,
	1981.
	
	\bibitem{QuaDRiGa_tech_rpt}
	S.~Jaeckel, L.~Raschkowski, K.~B\"{o}rner, L.~Thiele, F.~Burkhardt, and
	E.~Eberlein, ``{QuaDRiGa} - quasi deterministic radio channel generator user
	manual and documentation,'' Fraunhofer Heinrich Hertz Institute, Tech. Rep.
	v2.0.0, Aug. 2017.
	
	\bibitem{mirfarshbafan21}
	S.~H. {Mirfarshbafan}, S.~{Taner}, and C.~{Studer}, ``{SMUL-FFT:} a streaming
	multiplierless fast {Fourier} transform,'' \emph{{IEEE} Trans. Circuits Syst.
		{II}}, vol.~68, no.~5, pp. 1715--1719, Mar. 2021.
	
	\bibitem{sirpac21}
	O.~Casta{\~n}eda, Z.~Boynton, S.~H. Mirfarshbafan, S.~Huang, J.~C. Ye,
	A.~Molnar, and C.~Studer, ``A resolution-adaptive 8 $\text{mm}^2$ 9.98 {Gb/s}
	39.7 {pJ/b} 32-antenna all-digital spatial equalizer for {mmWave} massive
	{MU-MIMO} in 65nm {{CMOS}},'' in \emph{Proc. IEEE Eur. Solid-State Circuits
		Conf. (ESSCIRC)}, Sep. 2021.
	
	\bibitem{liu20}
	L.~{Liu}, G.~{Peng}, P.~{Wang}, S.~{Zhou}, Q.~{Wei}, S.~{Yin}, and S.~{Wei},
	``Energy- and area-efficient recursive-conjugate-gradient-based {MMSE}
	detector for massive {MIMO} systems,'' \emph{{IEEE} Trans. Signal Process.},
	vol.~68, pp. 573--588, Jan. 2020.
	
	\bibitem{tang21}
	W.~Tang, C.-H. Chen, and Z.~Zhang, ``A 0.58-mm2 {2.76-Gb/s 79.8-pJ/b 256-QAM}
	message-passing detector for a 128 × 32 massive {MIMO} uplink system,''
	\emph{{IEEE} J. Solid-State Circuits}, vol.~56, no.~6, pp. 1722--1731, Apr.
	2021.
	
	\bibitem{Rabaey_book}
	J.~M. Rabaey, A.~Chandrakasan, and B.~Nikoli\'c, \emph{Digital integrated
		circuits: A design perspective}.\hskip 1em plus 0.5em minus 0.4em\relax
	Prentice Hall, 2003.
	
	\bibitem{prabhu_phd}
	{Prabhu, Hemanth}, ``Hardware implementation of baseband processing for massive
	{MIMO},'' Ph.D. dissertation, {Lund University}, Sweden, 2017.
	
	\bibitem{Mahdavi19}
	M.~{Mahdavi}, O.~{Edfors}, V.~{{\"O}wall}, and L.~{Liu}, ``A {VLSI}
	implementation of angular-domain massive {MIMO} detection,'' in \emph{Proc.
		IEEE Int. Symp. Circuits and Syst. (ISCAS)}, May 2019, pp. 1--5.
	
	\bibitem{ZixiaoLi25}
	Z.~Li, S.~H. Mirfarshbafan, O.~Casta{\~n}eda, and C.~Studer, ``A
	deep-unfolding-optimized coordinate-descent data-detector {ASIC} for {mmWave}
	massive {MIMO},'' \emph{{IEEE} J. Sel. Areas Commun.}, Jan. 2025.
	
\end{thebibliography}
\end{document}